\documentclass[aps,pra,showpacs,twocolumn,superscriptaddress,nobibnotes,pdftex,pedfencoding=auto,colorlinks,citecolor=darkblue,linkcolor=darkred,urlcolor=darkblue]{revtex4-1}

\RequirePackage{subfigure}
\usepackage{lipsum}
\usepackage{graphicx}
\usepackage{color}
\usepackage{enumerate}
\usepackage{amsmath}
\usepackage{bm}
\usepackage{diagbox}
\usepackage{amssymb}
\usepackage{stmaryrd}
\usepackage{multirow}
\usepackage[normalem]{ulem}
\usepackage{float}
\usepackage{longtable,booktabs}
\usepackage[dvipsnames,table,xcdraw]{xcolor}
\usepackage[normalem]{ulem}
\usepackage{physics}
\usepackage{url}
\usepackage{soul} 
\usepackage[colorinlistoftodos]{todonotes}
\usepackage{verbatim}
\usepackage{comment}

\usepackage[draft]{pdfcomment}

\definecolor{darkblue}{rgb}{0.1,0.2,0.6} \definecolor{darkred}{rgb}{0.8,0.1,0.2}
\definecolor{darkgreen}{rgb}{0.0, 0.545098, 0.0}
\usepackage[normalem]{ulem}

\usepackage{siunitx}
\usepackage{cleveref}

\newcommand{\paren}[1]{\left(#1\right)}
\newcommand{\bracket}[1]{\left[#1\right]}

\def\beq{\begin{equation}}
\def\eeq{\end{equation}}	 

\long\def\/*#1*/{}

\begin{document}
\title{Framework for simulating gauge theories with dipolar spin systems}

\author{Di Luo}
\thanks{These authors contributed equally to this work.}
\affiliation{Department of Physics and IQUIST,  University of Illinois at Urbana-Champaign, IL 61801, USA} 
\affiliation{Institute for Condensed Matter Theory,  University of Illinois at Urbana-Champaign, IL 61801, USA}
\author{Jiayu Shen}
\thanks{These authors contributed equally to this work.}
\affiliation{Department of Physics and IQUIST,  University of Illinois at Urbana-Champaign, IL 61801, USA}
\author{Michael Highman}
\affiliation{Department of Physics and IQUIST,  University of Illinois at Urbana-Champaign, IL 61801, USA} 
\author{Bryan K.\ Clark}
\affiliation{Department of Physics and IQUIST,  University of Illinois at Urbana-Champaign, IL 61801, USA} 
\affiliation{Institute for Condensed Matter Theory,  University of Illinois at Urbana-Champaign, IL 61801, USA} 
\author{Brian DeMarco}
\affiliation{Department of Physics and IQUIST,  University of Illinois at Urbana-Champaign, IL 61801, USA} 
\author{Aida X.\ El-Khadra}
\affiliation{Department of Physics and IQUIST,  University of Illinois at Urbana-Champaign, IL 61801, USA} 
\author{Bryce Gadway}
\affiliation{Department of Physics and  IQUIST,  University of Illinois at Urbana-Champaign, IL 61801, USA} 

\begin{abstract}

Gauge theories appear broadly in physics, ranging from the Standard Model of particle physics to long-wavelength descriptions of topological systems in condensed matter.  
However, systems with sign problems  are largely inaccessible to classical computations and also beyond the current limitations of digital quantum hardware.
In this work, we develop an analog approach to simulating gauge theories with
an experimental setup that employs dipolar spins (molecules or Rydberg atoms).
We consider molecules fixed in space and interacting through dipole-dipole  interactions, avoiding the need for itinerant degrees of freedom.
Each molecule represents either a site or gauge degree of freedom, and Gauss's law is preserved by a direct and programmatic tuning of positions and internal state energies.
This approach can be regarded as a form of analog systems programming and charts a path forward for near-term quantum simulation. As a first step, we numerically validate this scheme in a small system study of U(1) quantum link models in (1+1) dimensions with link spin $S = 1/2$ and $S=1$ and illustrate
how dynamical phenomena such as string inversion and string breaking could be observed in near-term experiments. 
Our work brings together methods from atomic and molecular physics, condensed matter physics, high energy physics, and quantum information science for the study of nonperturbative processes in gauge theories.

\end{abstract}

\maketitle

\section{Introduction}

Gauge theories are fundamental to descriptions of a wide range of phenomena in many areas of physics. Euclidean lattice field theory~\cite{Wilson-Quarks-1974} has been developed into a general quantitative tool for the study of nonperturbative phenomena of gauge theories using Monte Carlo simulations on classical computers. In particular, the use of lattice gauge theory (LGT) to study quantum chromodynamics (QCD) has yielded profound insights into its nonperturbative dynamics and hence our understanding of particle and nuclear physics~\cite{Lattice-PhysTod,Kronfeld:2012uk,El-Khadra:2015lea,Ding:2015ona,Lattice-HeavyIon-Rev,Lehner:2019wvv,Aoki:2019cca}. However, problems involving nonequilibrium dynamics or systems described by complex actions suffer from the sign problem and are not easily amenable to Euclidean LGT simulations. Recent approaches based on tensor networks have been useful for investigating low-dimensional LGTs~\cite{LGTsQuantum-2019}, but their extension to higher dimensions may not be straightforward, due to the entanglement constraint.

The possibility of using quantum computations or simulations for these problems has led to renewed interest in Hamiltonian formulations~\cite{Kogut-Susskind-1975}. 
Quantum link models (QLMs)~\cite{Horn-QLM,Orland-QLM,Chandr-QLM}, an alternative formulation of LGTs, are well suited to studies of real-time dynamics on both classical and quantum devices. They use quantum spins in finite integer and half-integer representations of $S$ to replace the infinite gauge degrees of freedom. 

\begin{figure}[t]
  \centering
  \includegraphics[scale=.98]{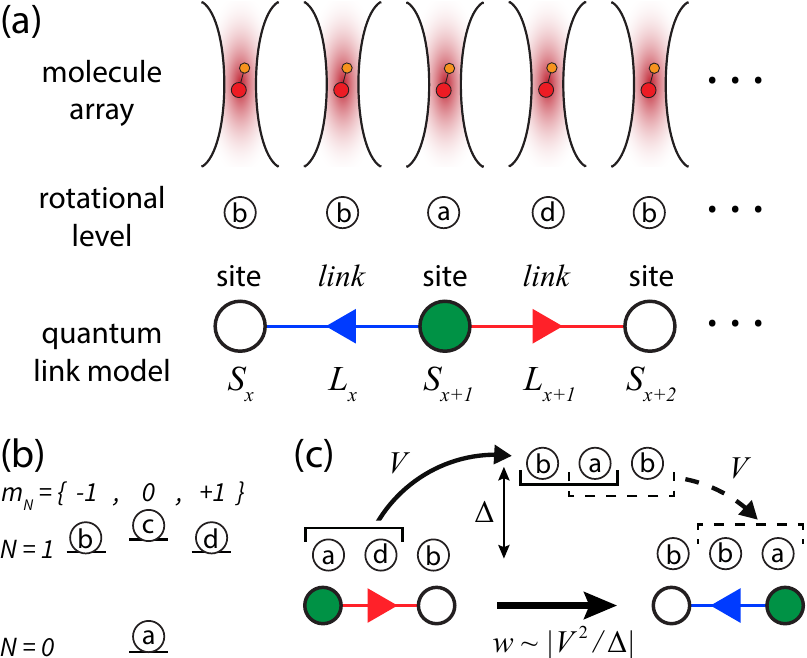}
  \caption{
  \textbf{Emulating quantum link models (QLMs) with arrays of dipolar molecules.}
  (a)~Mapping between the rotational levels of molecules in an array and the sites and links of the QLM for spin $S = 1/2$. The designation of particular molecules as sites or links ($S_x$, $L_x$, $S_{x+1}$, $L_{x+1}$ for a given unit cell) is enforced through local laser control of level-dependent light shifts. 
  (b)~Low-lying molecular rotational levels $|N,m_N\rangle$ and their redefinition in terms of states $\ket{a}$, $\ket{b}$, $\ket{c}$, and $\ket{d}$.
  (c)~The hopping of ``fermions'' between sites and the associated spin operations on the links are realized by a second-order dipolar exchange of rotational excitations.
  }
  \label{fig:IntroFig}
\end{figure}

Digital quantum computers, typically based on qubits and quantum circuits, are still in the noisy intermediate-scale 50
quantum (NISQ) era and are limited by both qubit number and gate depth due to noise and decoherence~\cite{PreskillNISQ}. Still, for small systems~\cite{Blatt-2016} and with the aid of variational techniques~\cite{Blatt-variational}, digital approaches to the quantum simulation of gauge theory dynamics have shown recent success. Alternatively, analog quantum simulators  --- special purpose quantum systems designed to implement the real-time evolution of model Hamiltonians --- may be more suitable for near-term investigations of LGT dynamics on moderate to large systems~\cite{Wiese-ReviewLGT-2013,Zohar-report-15,LGTsQuantum-2019,Axion-Bermudez,Kuno_2015,KUNO-gauge-higgs,Z2-Aid-theory,Z2-aid-expt,Z2-ess-expt,Jendrz-2019}.
While some problems are emulated naturally on physical platforms, such as realizing the Hubbard model with cold atoms in optical lattices~\cite{Jaksch98,Hofst-02} or realizing Heisenberg spin models with arrays of polar molecules or Rydberg atoms~\cite{Barnett-QM-polar-2006,Micheli-toolbox,Gorshkov-DipolarMagnetismPRA-2011,2015famr.book....3W,Browaeys-review}, matter-gauge dynamics is not. Here, we show that the physics of matter coupled to dynamical gauge fields can arise naturally in dipolar spin systems~\cite{Yan-DipDip-14,Hazzard-Gadway-PRL-14} if the elementary dipolar processes are restricted in a way that effectively imposes gauge invariance. Specifically, our approach relies on encoding the LGT Hamiltonian into a set of physically realizable degrees of freedom, for example in the internal states of polar molecules. This type of \emph{analog systems programming} or hardware-specific encoding is an intermediate approach between pure analog emulators, whose microscopic degrees of freedom closely match the emulated model, and fully digital simulations.

The encoding of matter-gauge dynamics into a pure spin model with nonitinerant particles addresses one of the main challenges facing the simulation of gauge theories based on atomic Hubbard models~\cite{Banerjee-GaugeFields-12,Banerjee-GaugeFields2-14}: the challenge of removing motional entropy and mitigating sources of heating~\cite{McKay_2011}. Such issues that plague itinerant systems are further compounded in schemes based upon spin- or species-dependent optical lattices~\cite{Banerjee-GaugeFields-12}, due to off-resonant light scattering~\cite{LeBlanc-species,McKay_2010}. These issues are avoided in pure spin systems, as motional entropy can be divorced from the dynamics of internal degrees of freedom~\cite{Hazzard-FarFrom} initialized with near-zero entropy. 

Recent theory work has shown that in experiments with Rydberg atom arrays~\cite{Bernien-nat}, LGT dynamics arises from Ising spin models by integrating out fermion fields~\cite{Surace-2019,Notarnicola-Rydberg-2020}.  The phenomenology of field theories, such as confinement, has also been demonstrated~\cite{StringSpin-2019,Monroe-Confine-2019} in other Ising spin systems.
The reduction of the required degrees of freedom by integration over fermion fields is a powerful approach, but it typically introduces projectors onto certain gauge field states that impose Gauss's law. For general link spin $S$, this can require unphysical forms of long-range interactions, and thus would be difficult to implement in analog simulations.
In our proposed approach we explicitly represent both the matter and gauge degrees of freedom with the matter-gauge dynamics arising due to dipole-mediated hopping of spin excitations in an array of nonitinerant dipolar spins.
Gauge invariance, namely, Gauss' law, is imposed in this construction by the application of local, state-dependent energy shifts that serve to constrain the dynamics of the spin excitations. 
As a first step, we specifically consider how a platform of trapped dipolar molecules with control of internal state energies can realize the analog simulation of a U(1) quantum link model in $(1+1)$ dimensions with spin $S = 1/2$ and $S = 1$. This approach relies on the dipolar nature of the spin-spin interactions, and could alternatively be realized in arrays of Rydberg atoms~\cite{Browaeys-review,Browaeys-opticalshift} or other dipolar systems~\cite{Cai-NV-dipolar}. 
We describe the detailed mapping between the states and parameters of the base molecular spin Hamiltonian and the target QLMs.
Our numerical simulations show that this scheme can allow for high fidelity analog simulations of dynamical phenomena fundamental to LGTs under realistic experimental conditions. 

\section{Quantum link models}
The local interaction terms of a gauge theory between matter and gauge-boson fields are imposed by the underlying gauge symmetry. In lattice gauge theory, the interaction terms involve matter fields at neighboring lattice sites. In order to preserve gauge invariance even at finite lattice spacing, the gauge bosons are usually represented by so-called link fields~\cite{Wilson-Quarks-1974}, which take continuous values being elements of a continuous gauge group. In the quantum link model version of LGTs, the link variables are instead represented by noncommuting, finite-dimensional operators that are analogous to quantum spin operators, 
a feature that makes QLMs more directly accessible to quantum simulation. Here, we consider QLMs of a U(1) LGT in $1+1$ dimensions in the Hamiltonian formulation with staggered fermions~\cite{Kogut-Susskind-1975}, which take the form
\begin{equation}
\begin{aligned}
H_\textrm{QLM} = & - w \sum _ { x } \left[ \psi _ { x } ^ { \dagger } U _ { x , x + 1 } \psi _ { x + 1 } + \psi _ { x + 1 } ^ { \dagger } U _ { x , x + 1 } ^ { \dagger } \psi _ { x } \right] \\
& + m \sum _ { x } ( - 1 ) ^ { x } \psi _ { x } ^ { \dagger } \psi _ { x } + \frac { g ^ { 2 } } { 2 } \sum _ { x } E _ { x , x + 1 } ^ { 2 }
,
\end{aligned}
\label{eq:QLM}
\end{equation}
where $x$ labels the spatial 
lattice sites, 
$\psi_x$ is the fermion operator with the staggered mass $m (-1)^x$, $w>0$ is the hopping parameter, $U_{x, x+1}$ is the link variable, $E_{x, x+1}$ is the electric flux for the U(1) gauge field on the link between $x$ and $x+1$, and $g$ is the gauge coupling~\cite{Kogut-Susskind-1975,Chandr-QLM}. 
In this paper we focus on QLMs with $S = 1/2$ and $S = 1$ representations for the link variables, but this may be generalized to larger $S$. Note that the 
continuous gauge symmetry
is recovered in the $S \rightarrow \infty$ limit~\cite{Banerjee-GaugeFields-12}. The physical Hilbert space of the \hyperref[eq:QLM]{QLM} is constrained by the gauge symmetry through Gauss's law and the gauge transformations are generated by the the Gauss-law operator
$\widetilde { G } _ { x } = \psi _ { x } ^ { \dagger } \psi _ { x } - E _ { x , x + 1 } + E _ { x - 1 , x } + \frac { 1 } { 2 } [ ( - 1 ) ^ { x } - 1 ]$, where the last term stems from using staggered fermions.

\section{Dipolar molecules}
A dipolar molecule is effectively a quantum rotor with angular momentum and projection eigenstates $\ket{N_\alpha, m_{N_\alpha}}$. 
We restrict our consideration to electronic, vibrational, and hyperfine ground states in the absence of large dc electric fields. Furthermore, we only consider rotational angular momentum states with $N_{\alpha} \in \{0,1\}$, hereafter using the notation $\ket{a} \equiv \ket{0, 0}$, $\ket{b} \equiv \ket{1, -1}$, $\ket{c} \equiv \ket{1, 0}$ and $\ket{d} \equiv \ket{1, 1}$ for the states we consider (see Fig.~\ref{fig:IntroFig}). 
The Hamiltonian for a system of fixed dipolar molecules (\hyperref[eq:dipole-mol-ham]{DMH}) is
\begin{equation}
\begin{aligned}
    H_\textrm{DMH} = & 
    \sum_{i, \alpha} \left( 2 h B_{\mathrm{rot}} N_{\alpha} (N_{\alpha} + 1) + \epsilon_{i, \alpha} \right) b^\dagger_{i, \alpha} b_{i, \alpha} \\ 
    & + \frac{1}{2} \sum_{i, j} \sum_{\alpha, \beta, \gamma, \eta} V_{i, j}^{\alpha, \beta; \gamma, \eta} b^\dagger_{i, \gamma} b^\dagger_{j, \eta} b_{j, \beta} b_{i, \alpha}
    .
\end{aligned}
\label{eq:dipole-mol-ham}
\end{equation}
where $b^\dagger$ and $b$ are hard-core bosonic creation and annihilation operators, $B_\textrm{rot}$ is the molecule's rotational constant, $i,j$ label molecular positions, and $\alpha,\beta,\gamma,\eta$ label angular momentum states. The $\epsilon_{i, \alpha}$ are the additional position- and/or state-dependent contributions to the single-molecule energies, which 
include both spatially uniform but state-dependent terms arising from weak electric fields or nuclear-rotational coupling~\cite{Gorshkov-DipolarMagnetismPRA-2011}, as well as position-dependent and state-dependent energy shifts, which can be engineered through local differential ac Stark shifts owing to the anisotropic polarizability of molecules~\cite{Neyenhuis-Polarizability-12}. In particular, local control over molecular light shifts could be naturally incorporated in molecule arrays with individual laser addressing~\cite{Anderegg1156}. The second line of Eq.~(\ref{eq:dipole-mol-ham}) represents the dipole-dipole interaction $V_{i, j}^{\alpha,\beta;\gamma,\eta}$, where 
$\alpha,\beta;\gamma,\eta$ label pairs of initial and final angular momentum states. The forms of the interaction terms are naturally restricted by the dipole selection rules $\Delta N = \pm 1$ and $\Delta m_N = 0, \pm 1$, but importantly their dipolar nature allows for the populations of the various molecular levels to be nonconserved (i.e., dipolar interactions allow for the interconversion of rotational and orbital angular momentum~\cite{EdH}). Experimentally, one can use laser power and polarization to tune the relative $\epsilon_{i,\alpha}$ terms, and we consider control of the $V_{i, j}^{\alpha,\beta;\gamma,\delta}$ magnitudes through control of intermolecular distances.

\begin{table}[t!]
\begin{center}
\begin{tabular}{c c | c c}
	\hline\hline
	$S = 1/2$ & & $S = 1$ & \\
	\hline
	\begin{tabular}{@{}c@{}}Molecular \\ Levels\end{tabular} & \begin{tabular}{@{}c@{}}QLM \\ States\end{tabular} & \begin{tabular}{@{}c@{}}Molecular \\ Levels\end{tabular} & \begin{tabular}{@{}c@{}}QLM \\ States\end{tabular} \\
	\hline
	$\ket{a}_{S}$ & Occupied & $\ket{a}_{S}$ & Occupied \\
	$\ket{b}_{S}$ & Unoccupied & $\ket{c}_{S}$ & Unoccupied \\
	\hline
	$\ket{b}_{L}$ & $S^3 = -1/2$ & $\ket{d}_{L}$ & $S^3 = -1$ \\
	$\ket{d}_{L}$ & $S^3 = 1/2$ & $\ket{b}_{L}$ & $S^3 = 0$ \\
	& & $\ket{c}_{L}$ & $S^3 = 1$ \\
	\hline\hline
\end{tabular}
\end{center}
\caption{Mapping from molecule levels to QLM states.}
\label{table:map}
\end{table}

\section{Effective Hamiltonian}
We now generate a map between the parameters of the \hyperref[eq:QLM]{QLM} and the physical parameters of the \hyperref[eq:dipole-mol-ham]{DMH}. In our mapping, every site and link in the QLM maps to a different individual molecule; the Hilbert spaces are mapped as in Table~\ref{table:map}. In this construction, the rotational levels on the ``site'' molecules are used to represent the  ``fermions'' (which are hard core bosons in one dimension) while rotational levels on the ``link'' molecules represent the link gauge fields.
For a QLM with staggered fermions, each unit cell 
corresponds to two sites (odd $x$ and even $x$) and two links, with the link between sites $S_x$ and $S_{x+1}$ labeled as $L_x$. 

Letting $H_0$ be the one-body terms in the first line of Eq.~(\ref{eq:dipole-mol-ham}), we tune 
the energies $\epsilon_{i,\alpha}$ so that Gauss's law 
is satisfied in the QLM and 
so that all of the dipolar configurations satisfying Gauss's law are nearly degenerate in $H_0$ while all of the other configurations are separated 
from the Gauss-law configurations by an energy scale $\Delta \gg V$.
While we keep $m,g^2 \ll V \ll \Delta$ to preserve this condition, this still allows for the tuning of $m$ and $g^2/2$ on scales comparable to the hopping $w$.
If the molecular system is prepared s
in an initial state that satisfies Gauss's law, energy constraints will ensure that the time-evolved state 
will remain in the physical Hilbert space.

The hopping term $- w \sum _ { x } [ \psi _ { x } ^ { \dagger } U _ { x , x + 1 } \psi _ { x + 1 } + \text{H.c.}]$ of the \hyperref[eq:QLM]{QLM} involves two sites and one link, and implies changes in the states of three molecules, $S_x$, $L_x$, and $S_{x+1}$.
Because the \hyperref[eq:dipole-mol-ham]{\hyperref[eq:dipole-mol-ham]{DMH}}  [Eq.~(\ref{eq:dipole-mol-ham})] contains only two-body interactions, matching to the
\hyperref[eq:QLM]{QLM} hopping term proceeds by constructing 
the quasidegenerate effective Hamiltonian to second order \cite{winkler2003spin} and includes a combination of exchange terms $V_{i,j}$ and state-dependent energies $\epsilon_{i,\alpha}$. 
In the following two sections we describe the details of this mapping for $S=1/2$ and $S=1$.

\begin{figure}[b]
  \centering
  \includegraphics[width=1.0\columnwidth]{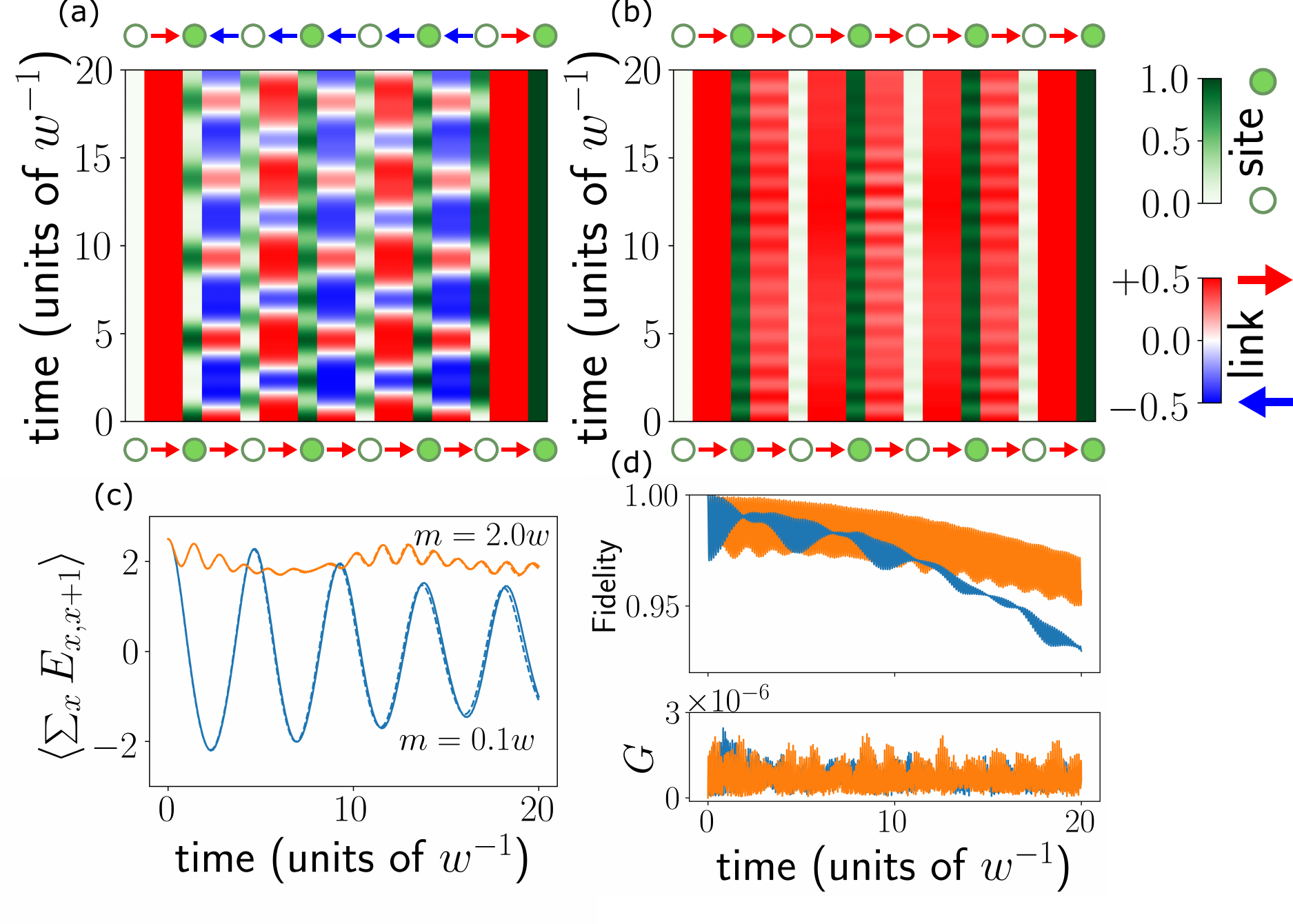}
  \caption{
  Real-time evolution of densities of sites and links in the dipolar molecular system to simulate the $S=1/2$ QLM on three unit cells. The dynamics for initialized strings of right-pointing electric fields are shown for the cases of small mass, (a) $m = 0.1 w$, and large mass, (b) $m = 2.0 w$. Time is in units of the inverse hopping, $w^{-1}$. For small mass (a), the electric field of the string undergoes large-scale oscillations. For large mass (b), the string stays roughly fixed, with only small fluctuations of the charge densities and link spins.
  To note for both (a) and (b), the outermost sites and links are fixed because of the open boundary conditions.
  (c)~Electric fluxes summed over all dynamical links for $m = 0.1 w$ (blue) and $m = 2.0 w$ (orange). Solid and dashed lines relate to the \hyperref[eq:dipole-mol-ham]{DMH} and \hyperref[eq:QLM]{QLM} dynamics, respectively.
  (d) Top: Fidelity of the dipolar molecular wavefunction versus the QLM wavefunction. 
  Bottom: The effective gauge invariance parameter $G \equiv \sum_x | \langle  \tilde{G}_x \rangle | / L$ \cite{Banerjee-GaugeFields-12} at the two mass values shown in (a) and (b).
  }
  \label{fig:four_panel_S05}
\end{figure}

\section{Realization and tests of $S = 1/2$ QLM}
Here we describe details specific to the mapping for the case $S=1/2$ and we numerically validate this mapping. In this mapping, the molecular state $\ket{c}$ is kept energetically decoupled from gauge-invariant initial states, which can be achieved by means of a small dc electric field. Furthermore, local light shifts can be used to decouple the molecular state $\ket{d}$ from dynamics at the site positions. On the links, while the molecular state $\ket{a}$ does not map directly to anything in the \hyperref[eq:QLM]{QLM} Hilbert space, it is utilized to help mediate the second-order process needed to describe the
fermion hopping interaction in the \hyperref[eq:QLM]{QLM}.

A three-body hopping process is illustrated in Fig.~\ref{fig:IntroFig} for the example: 
$\ket{a}_{S_x}\ket{d}_{L_x}\ket{b}_{S_{x+1}} \rightarrow \ket{b}_{S_x}\ket{b}_{L_x}\ket{a}_{S_{x+1}}$. 
In the dipole system this is a second-order process, which can proceed via 
$\ket{a}_{S_{x}}\ket{d}_{L_x}\ket{b}_{S_{x+1}} \xrightarrow[]{\text{virtual}}\ket{b}_{S_x}\ket{a}_{L_x}\ket{b}_{S_{x+1}} \xrightarrow[]{\text{virtual}} \ket{b}_{S_x}\ket{b}_{L_x}\ket{a}_{S_{x+1}}$, where $\ket{b}_{S_x}\ket{a}_{L_x}\ket{b}_{S_{x+1}}$ is an intermediate state outside the physical Hilbert space 
with an energy difference $\Delta \gg V$.  

The second-order hopping process has a term given by
\begin{equation}
     - w = \frac { 1 } { 2 } V_{S_x, L_x}^{b, a; a, d} V_{L_x, S_{x+1}}^{b, a; a, b} \left[ \frac{1}{\Delta \epsilon_{1,x}} + \frac{1}{\Delta \epsilon_{2,x}} \right] \ ,
\label{eq:hopping_1}
\end{equation}
with $\Delta \epsilon_{1,x} = \epsilon_{S_x, a} + \epsilon_{L_x, d} - \epsilon_{S_x, b} - \epsilon_{L_x, a}$ and $\Delta \epsilon_{2,x} = \epsilon_{L_x, b} + \epsilon_{S_{x+1}, a} - \epsilon_{L_x, a} - \epsilon_{S_{x+1}, b}$. This yields 
one equation for every $x$. 
While the QLM has a two-site unit cell that repeats, the microscopic parameters of the underlying \hyperref[eq:dipole-mol-ham]{DMH} can be varied slightly between $x$ and $x+2$, e.g., to allow for the mitigation of undesired processes resulting from the long-ranged dipolar interactions. Given a target $w$ value, we can then simultaneously solve for all these equations up to corrections of order greater than $O ( V^2 / \Delta)$, which generates relationships between the various $V$ and $\epsilon$ values. These parameters are chosen so as to preserve Gauss's law, and furthermore so that they are physically reasonable. 
Finally, in addition to the kinetic terms that appear in the second-order effective model, second-order self-interaction terms appear as well. 
These small diagonal terms can be fully compensated by a slight renormalization of the \hyperref[eq:dipole-mol-ham]{DMH} energy terms $\epsilon_{i,\alpha}$~(see Tables~\ref{table:energy_simulation_S05} and \ref{table:energy_simulation_S1}).

We numerically confirm the mapping for $S=1/2$ by simulating the \hyperref[eq:QLM]{QLM} on three unit cells with open boundary conditions, and comparing to the full simulation of the \hyperref[eq:dipole-mol-ham]{DMH}. Figures \ref{fig:four_panel_S05}(a)-\ref{fig:four_panel_S05}(c) show the \hyperref[eq:dipole-mol-ham]{DMH} dynamics of the sites and links, for an initial product state configuration with staggered site occupations and polarized electric fields. Note that the electric flux energy is always a constant in the $S=1/2$ QLM, and therefore $g^2$ is irrelevant.
For a small mass, $m = 0.1 w$, the dynamics of the \hyperref[eq:dipole-mol-ham]{DMH} reveal a string inversion of the electric fluxes [Figs.~\ref{fig:four_panel_S05}(a) and~\ref{fig:four_panel_S05}(c)]. For a large mass, $m = 2.0 w$ (more than three times the reported critical mass $m_c = 0.655w$~\cite{PhysRevLett.112.201601}), the system shows little dynamics, as there is almost a static flux string with small fluctuations [Figs.~\ref{fig:four_panel_S05}(b) and~\ref{fig:four_panel_S05}(c)].

We additionally compute the wave-function fidelity and violation of Gauss's law, shown in Fig.~\ref{fig:four_panel_S05}(d). The fidelity is defined as $|\langle \psi_\mathrm{QLM} |  \psi_\mathrm{DMH} \rangle |^2$, where $\psi_\mathrm{DMH}$ is the wave-function from the \hyperref[eq:dipole-mol-ham]{DMH} and $\psi_\mathrm{QLM}$ is the wave-function mapped from the \hyperref[eq:QLM]{QLM} into the \hyperref[eq:dipole-mol-ham]{DMH} space. The fidelity of our scheme is greater than 0.9 out to $t=20 w^{-1}$, and Gauss's law is preserved over this time range at the $10^{-6}$ level. For the molecule NaRb (body-frame dipole moment of 3.3~D) and a minimum separation of 0.5~$\mu$m, these calculations relate to a hopping rate of $w/h = 41.3$~Hz (with $h$ Planck's constant). This robust energy scale should be compatible with long molecule trapping times~\cite{Chotia-PRL} and coherence times~\cite{Yan-DipDip-14,Hazzard-Gadway-PRL-14}. In addition, larger values of the hopping can be achieved by reducing the scale of imposed energy penalties $\Delta$,
albeit with fidelities lower than those shown in Fig.~\ref{fig:four_panel_S05}(d).

\begin{figure}[b]
  \centering
    \includegraphics[width=1.0\columnwidth]{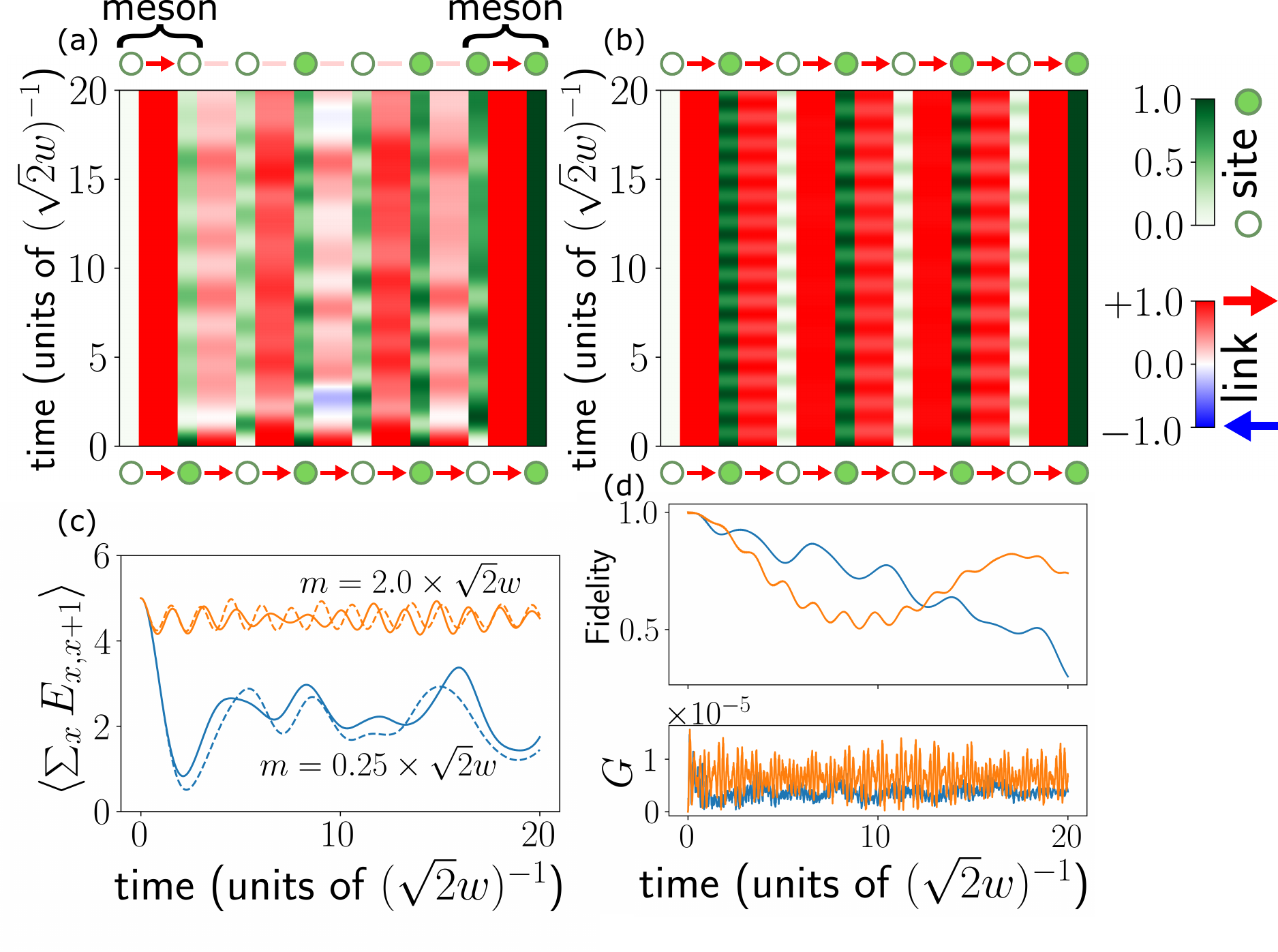}
  \caption{
  Real-time evolution of fermions and links in the dipolar molecular system to simulate the $S=1$ QLM on three unit cells with $g^2 = \sqrt{2} w$ starting from a string of right-pointing electric fields at (a) $m = 0.25 \times \sqrt{2} w$ and (b) $m = 2.0 \times \sqrt{2} w$. Time is in units of $(\sqrt{2} w)^{-1}$. With a small mass (a), the string breaks (modulo finite size effects~\cite{Pichler-PRX}) reaching values close to zero on the hopping timescale, resulting in two approximate mesons on the edges and approximate vacuum in between. With a large mass (b), the string approximately remains, with small fluctuations in densities.
  In (a) and (b), the densities of the two sites and two links on the edges are fixed due to the open boundary condition.
  (c)~Electric fluxes summed over all dynamical links for both $m = 0.25 \times \sqrt{2} w$ (blue) and $m = 2.0 \times \sqrt{2} w$ (orange). Solid and dashed lines relate to the \hyperref[eq:dipole-mol-ham]{DMH} and \hyperref[eq:QLM]{QLM} dynamics, respectively. (d)~Top: Fidelity of the dipolar molecular wavefunction versus the QLM wavefunction.
  Bottom: The effective gauge invariance parameter $G \equiv \sum_x | \langle  \tilde{G}_x \rangle | / L$~\cite{Banerjee-GaugeFields-12} at two values of masses in (a) and (b).
  }
  \label{fig:four_panel_S1}
\end{figure}

\section{Realization and tests of $S = 1$ QLM}
The $S=1$ case is realized similarly to $S=1/2$; however more internal levels are used to represent the larger number of link spin values. The link states are represented by the various $N=1$ rotational sublevels as in Table~\ref{table:map}. For the sites, the $\ket{b}$ and $\ket{d}$ levels are decoupled from the dynamics by large local light shifts. Unlike for $S=1/2$, in the $S=1$ case 
second-order self-interaction processes, such as $\ket{c}_{S_x}\ket{b}_{L_x}\ket{a}_{S_{x+1}} \xrightarrow[]{\text{virtual}}\ket{c}_{S_x}\ket{a}_{L_x}\ket{c}_{S_{x+1}} \xrightarrow[]{\text{virtual}} \ket{c}_{S_x}\ket{b}_{L_x}\ket{a}_{S_{x+1}}$, 
cannot be entirely removed through coordination of the $\epsilon_{i,\alpha}$ terms. This comes from the fact that they are not simply renormalized one-body terms. For example, the above process generates an $O ( V^2 / \Delta)$ term of the form $b_{L_x, b}^\dagger b_{L_x, b} b_{S_{x+1}, a}^\dagger b_{S_{x+1}, a}$ that is not in the \hyperref[eq:QLM]{QLM}. Nevertheless, 
these additional terms are diagonal in the molecular $\lbrace \ket{N, m_N} \rbrace$ basis and still preserve Gauss' law, but cannot be removed because the simultaneous constraints for 
$w$ and setting these ``extra'' terms to zero results in an overdetermined system of equations. To overcome this, we introduce a new length scale.  
 
For every $x$, we set the distances between molecules representing $S_x$ and $L_x$ to be small and denote the characteristic scale of the dipole-dipole interaction between them as $V_{\mathrm{short}}$. Meanwhile, 
the distances between molecules representing $L_x$ and $S_{x+1}$ are chosen to be larger with a 
characteristic scale for the dipole-dipole interaction between them of $V_{\mathrm{long}}$, and $V_{\mathrm{long}} \ll V_{\mathrm{short}}$. The hopping parameter $w$ is $O (V_{\mathrm{short}} V_{\mathrm{long}} / \Delta)$. The second-order self-interaction term for $S_x$ and $L_x$ is $O (V_{\mathrm{short}}^2 / \Delta)$ and that for $L_x$ and $S_{x+1}$ is $O (V_{\mathrm{long}}^2 / \Delta)$, which 
can be safely neglected. Therefore, we only need to consider the equations between $S_x$ and $L_x$, which decrease the total number of equations and leave them underdetermined. A nonunique solution to these equations can then be found,
and we can obtain experimental parameters such that these second-order terms are made small or can be removed from the effective Hamiltonian. To note, for larger $V_{\mathrm{short}} / V_{\mathrm{long}}$ ratios, higher-order terms eventually limit this minimization of the extra terms.
We set the intermolecular distances as $r_{S_{2n+1}, L_{2n+1}}=r_{S_1, L_1}$, $r_{L_{2n+1}, S_{2n+2}} = \gamma r_{S_1, L_1}$, $r_{S_{2n+2}, L_{2n+2}}= \beta r_{S_1, L_1}$,  and $r_{L_{2n+2}; S_{2n+3}} = \beta \gamma r_{S_1, L_1}$ for every $n$, where $n$ labels the unit cell. $\gamma$ is defined as a variable long-short distance ratio greater than or equal to one. $\beta$ is fixed by energy conditions and independent of $\gamma$.

To benchmark our scheme for $S=1$, we perform exact diagonalization on three unit cells with open boundary conditions. The initial configuration is chosen as shown in Fig.~\ref{fig:four_panel_S1}, relating to a flux string connecting static charges. String breaking is a key dynamical phenomenon in high-energy physics, found in QCD~\cite{stringB} as well as the simpler Schwinger model~\cite{StringSchwing}. Depending on the parameters of the $S=1$ QLM, the initial configuration as a string may break into approximate vacuum in the middle region, resulting in the production of two mesons on the edge. To verify that our scheme reflects the physical dynamics properly, we investigate both a small-mass scenario [Fig.~\ref{fig:four_panel_S1}(a), \{$m = 0.25 \times \sqrt{2} w, g^2 = \sqrt{2} w$\}] and a large-mass scenario [Fig.~\ref{fig:four_panel_S1}(b), \{$m = 2.0 \times \sqrt{2} w, g^2 = \sqrt{2} w$\}], which should result in string breaking and a stabilized string, respectively. Figure ~\ref{fig:four_panel_S1} shows that the dynamics of the dipolar molecular system (solid lines) reflects this behavior. We find good agreement with the expected dynamics of the target \hyperref[eq:QLM]{QLM} (dashed line). Specifically, for large mass, the string stays approximately in its initial configuration up to small fluctuations, while the string of the small mass case breaks on the hopping timescale. These results are consistent with the estimated critical length $L_c = 4 m / g^2 + 3$~\cite{Banerjee-GaugeFields-12}.

\begin{figure}[b]
  \centering
  \includegraphics[width=0.85\linewidth]{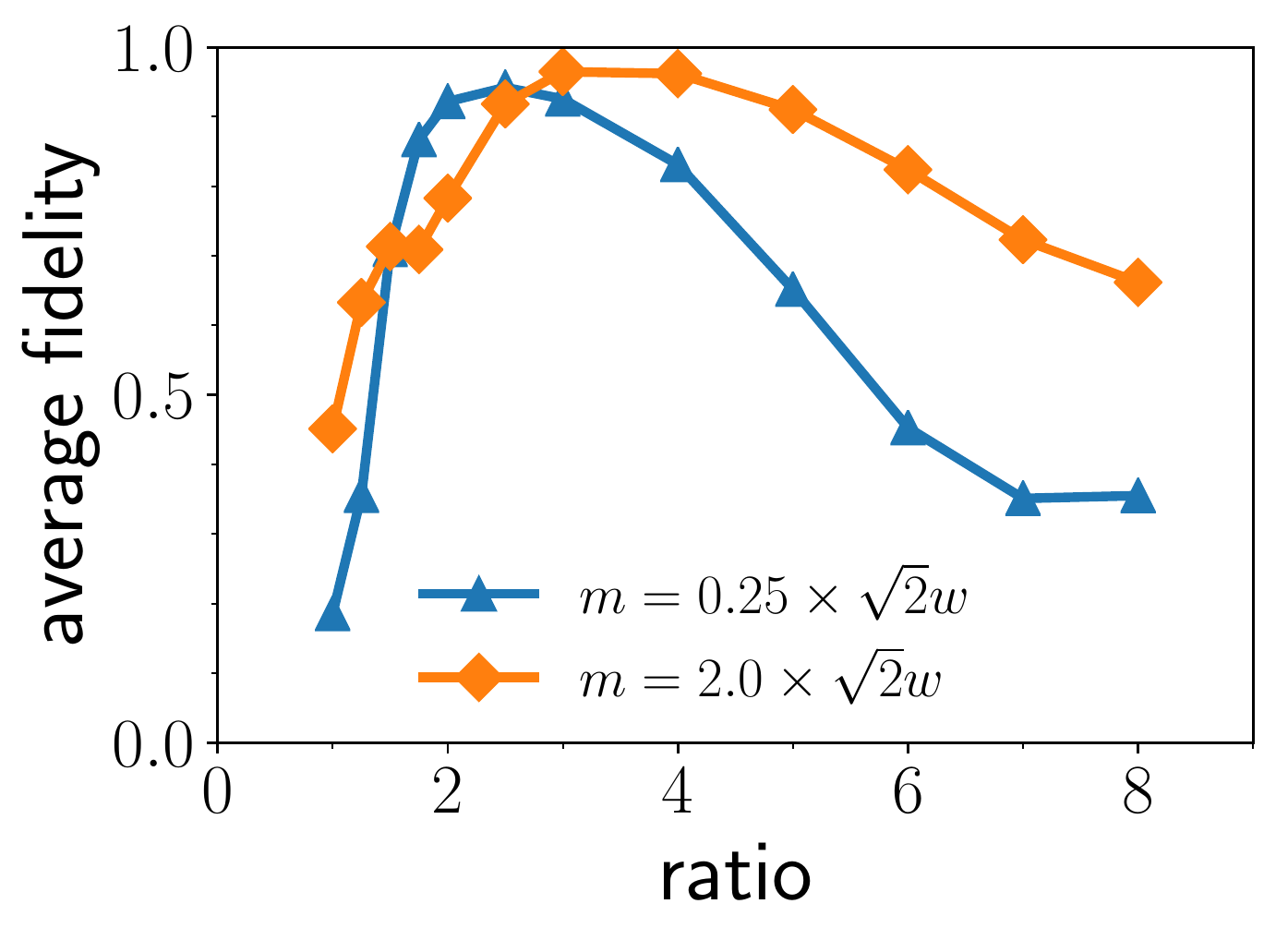}
  \caption{
  Average fidelity of the calculated DMH dynamics based on the mapping to the $S=1$ QLM, plotted versus the ``long-short'' distance ratio $\gamma$ used to mitigate the influence of ``extra'' gauge-invariant terms. The plotted curves are for mass values of $m = 0.25 \times \sqrt{2} w$ and $m = 2.0 \times \sqrt{2} w$. These represent the fidelity when starting from the initial product state as in Fig.~\ref{fig:four_panel_S1}, comparing the DMH-evolved state to the evolution under the ideal QLM, with averaging over the time period $0\leq t \leq 20$ [with time in units of $(\sqrt{2} w)^{-1}$]. An initial rise in fidelity is seen for moderate ratios as the ``extra'' gauge invariant terms are suppressed, but it decreases for larger ratios as higher-order terms become important.
  }
  \label{fig:multi_ratio_average_fidelity}
\end{figure}

Gauss's law is preserved to high accuracy [see Fig.~\ref{fig:four_panel_S1}(d)]; however, the fidelity of the \hyperref[eq:dipole-mol-ham]{DMH} ``simulator'' drops from 1.0 to roughly 0.5 over 10--20 time units [in terms of $(\sqrt{2} w)^{-1}$]. This infidelity comes almost entirely from the additional gauge-invariant self-interaction terms that do not appear in the \hyperref[eq:QLM]{QLM}. These terms slightly modify the frequencies of oscillations in the \hyperref[eq:dipole-mol-ham]{DMH} and \hyperref[eq:QLM]{QLM} dynamics, and thus have a large influence on the fidelity at long times, but otherwise do not alter the expected QLM phenomenology [see Fig.~\ref{fig:four_panel_S1}(c)].

While one approach to improving this fidelity is simply adding the additional gauge-invariant terms to the target \hyperref[eq:QLM]{QLM} (as is done in Ref.~\cite{Banerjee-GaugeFields-12}), we can suppress the additional gauge invariant terms and improve the fidelity by adjusting the ``long-short'' distance ratios.  In this work we use the value of $V_{\text{short}} / V_{\text{long}} = 1.5^3 \approx 3.4$ (see Fig.~\ref{fig:four_panel_S1}), which represents a nonoptimal but more experimentally realistic compromise;
the parameters used in Fig.~\ref{fig:four_panel_S1} already relate to hopping energies of $\sqrt{2} w/h = 3.2$~Hz for NaRb (3.3~D) and an assumed minimum spacing of 0.5~$\mu$m.  

\begin{figure}[b]
  \centering
  \includegraphics[width=1.0\linewidth]{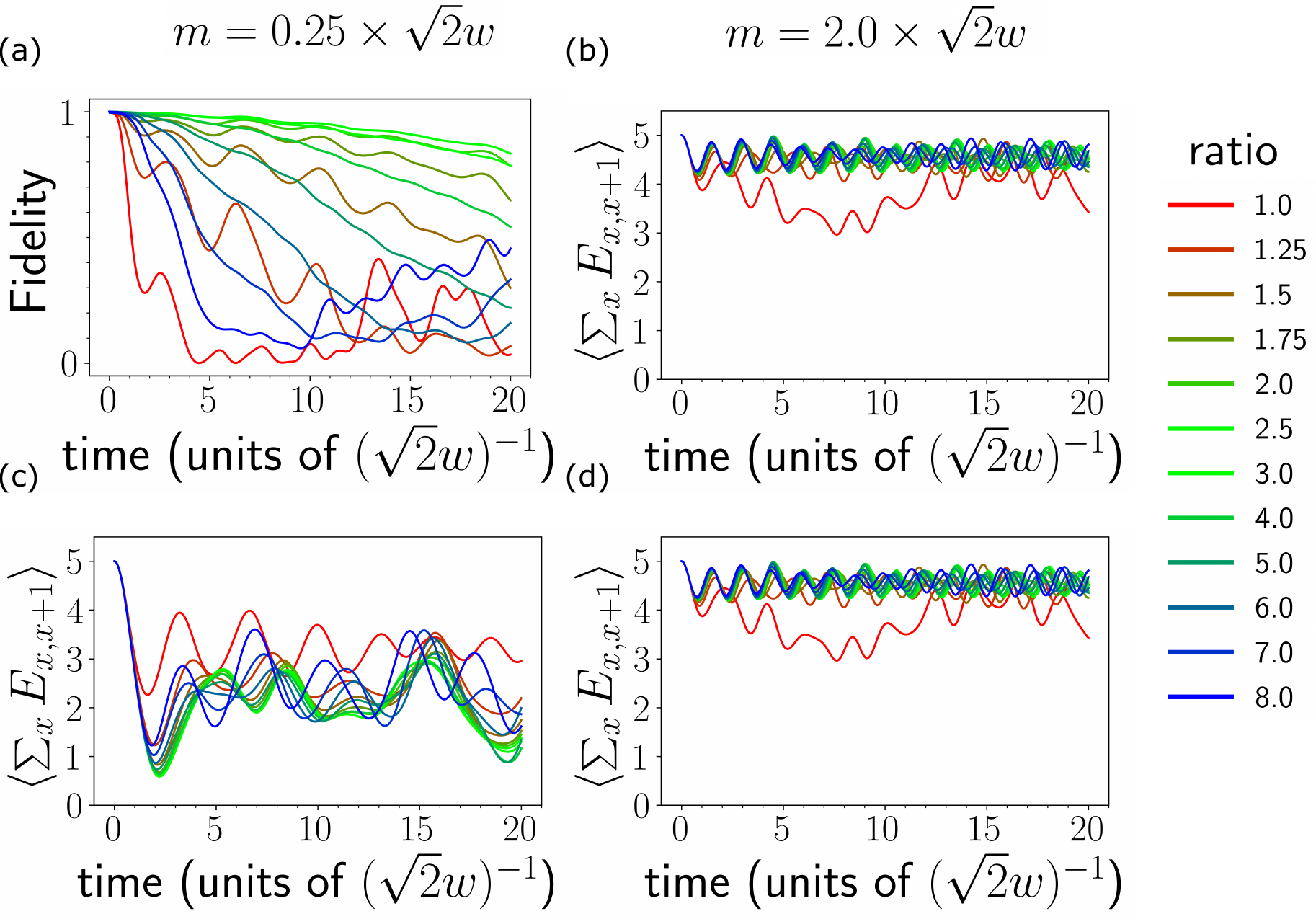}
  \caption{Effect of additional gauge-invariant terms generated by the dipolar molecular system for simulation of the $S=1$ QLM. This figure considers the influence of these terms on the fidelity (when comparing the DMH simulations to the ideal QLM) and the dynamics of key physical observables.
  Fidelity versus time for (a) $m = 0.25 \times \sqrt{2} w$ and (b) $m = 2.0 \times \sqrt{2} w$ with various long-short distance ratios.
  The sum of electric fields for (c) $m = 0.25 \times \sqrt{2} w$ and (d) $m = 2.0 \times \sqrt{2} w$ with various long-short distance ratios.
  For all panels, time is in units of $(\sqrt{2} w)^{-1}$.
  While the fidelity (as compared to the ideal QLM) is strongly dependent on the ratio value, the physical observations of string breaking in (c) and a fixed string with small fluctuations in (d) are robust over a very large range of ratio values.
  }
  \label{fig:multi_ratio_fidelity_efield}
\end{figure}

We have investigated the effect of the long-short distance ratio $\gamma$ on fidelity. Figures \ref{fig:multi_ratio_average_fidelity} and \ref{fig:multi_ratio_fidelity_efield} demonstrate the average fidelity up to time $t = 20 (\sqrt{2} w)^{-1}$ at various long-short distance ratios for $m = 0.25 \times \sqrt{2} w$ and $m = 2.0 \times \sqrt{2} w$. It is shown that the fidelity is low for both small and large $\gamma$, while the optimal value is achieved at an intermediate ratio between 2 and 3. For small long-short distance ratios, certain second-order self-interactions at the order $O ( V_{\mathrm{long}}^2 / \Delta)$ are not sufficiently suppressed, which results in a low fidelity. For large long-short distance ratios, higher-order self-interaction terms [such as fourth-order self-interactions, i.e., at the order $O (V_{\mathrm{short}}^4 / \Delta^3)$, for the pair of molecules $S_x$ and $L_x$ separated by a short distance] may become comparable to the hopping parameter $w \sim O (V_{\mathrm{short}} V_{\mathrm{long}}/ \Delta)$ and are thus not negligible.
Therefore, the optimal choice of $\gamma$ for fidelity should balance the trade-off between suppressing second-order self-interaction terms at the order $O ( V_{\mathrm{long}}^2 / \Delta)$ and avoiding high-order self-interactions, which is attained in the intermediate ratio. The effect of long-short distance ratios is further presented in Figs.~\ref{fig:multi_ratio_fidelity_efield}(a) and \ref{fig:multi_ratio_fidelity_efield}(b) for real-time fidelity changes.

The long-short distance ratios that provide the highest fidelity (as compared to the target QLM) may be less ideal from a practical perspective, as they result in lower hopping parameters $w$ for a fixed minimum separation of the molecules.
However, the reduction of the fidelity due to the extra gauge-invariant terms in the $S=1$ case does not necessarily preclude the DMH dynamics from displaying the physical processes of interest.
In our case, at a suboptimal long-short distance ratio, even though the fidelity may not be very high, the observed phenomenology is not significantly altered by the extra gauge invariant terms. In Figs.~\ref{fig:multi_ratio_fidelity_efield}(c) and \ref{fig:multi_ratio_fidelity_efield}(d), we compare the sum of electric fluxes, an indicator of string breaking, over various long-short distance ratios. Although different long-short distance ratios 
have very different fidelities, 
they result in 
similar string breaking phenomena except for $\gamma = 1.0$. Qualitatively, at all of the ratios except for $\gamma = 1.0$, our results at $m = 0.25 \times \sqrt{2} w$ reveal string breaking while the results at $m = 2.0 \times \sqrt{2} w$ do not. This provides some evidence that we may be able to choose suboptimal (in terms of fidelity with respect to the ideal QLM) but experimentally favorable parameters to explore the physics of interest through analog simulation.

\section{Conclusions and outlook}
In this paper, we propose an approach for simulating 
quantum link models based on the restricted internal-state dynamics of fixed dipolar spins.
Our numerical tests of simple U(1) LGTs in $1+1$ dimensions show that this approach enables the experimental exploration of important dynamical phenomena such as string inversion and string breaking.
Further directions are suggested by the present work, including extensions to realizations of higher $S$ values, non-Abelian LGTs, quantum mechanical $\theta$ angles~\cite{ThetaAngle}, and QLM dynamics in higher dimensions~\cite{2D-QLMs}. Moreover, while our scheme based on dipolar spin interactions is of relevance to a range of systems, such as cold molecules and Rydberg atom arrays, the potential reach could be greatly broadened by generalizing this framework to generic spin systems, as realized by most physical quantum information platforms.

\section*{Acknowledgements}
We thank Patrick Draper, Yannick Meurice, Jesse Stryker, Ming Li, and Svetlana Kotochigova for discussions. This work was supported in part by the U.S. Department of Energy under Award No.~DE-SC0019213 (D.L., J.S., B.K.C., B.D., A.X.K., and B.G.) and by the National Science Foundation Graduate Research Fellowship Program under Grant No.~DGE–1746047 (M.H.).

Note added --- We note that a recent experiment has reported the analog simulation of $S=1/2$ QLMs with scalar neutral atoms in an optical superlattice~\cite{yang2020observation}.

\appendix

\section{\label{sec:sm_qlm}Quantum link models}

We start with the Hamiltonian for $(1+1)$-dimensional U(1) LGT with staggered fermions in the temporal gauge, 
\begin{equation}
\begin{aligned}
H_\textrm{LGT} & = - w \sum _ { x } \left[ \psi _ { x } ^ { \dagger } U _ { x , x + 1 } \psi _ { x + 1 } + \psi _ { x + 1 } ^ { \dagger } U _ { x , x + 1 } ^ { \dagger } \psi _ { x } \right] \\ & + m \sum _ { x } ( - 1 ) ^ { x } \psi _ { x } ^ { \dagger } \psi _ { x } + \frac { g ^ { 2 } } { 2 } \sum _ { x } E _ { x , x + 1 } ^ { 2 }
.
\end{aligned}
\label{eq:sm_qlm}
\end{equation}
The link variables $U_{x, x+1} = \exp (i a g A_{x, x+1})$ take continuous values in the group U(1), where $A_{x, x+1}$ is the spatial component of the U(1) gauge field and $a$ is the lattice spacing. The electric flux $E_{x, x+1} = - i \frac{1}{ag} \frac{\partial}{\partial A_{x, x+1}}$ is proportional to the canonical momentum of $A_{x, x+1}$ and can take any integer values. Commutation relations for quantum operators on a link are $[ U _ { x , x + 1 }, U _ { x , x + 1 } ^ { \dagger } ] = 0$, $[ E_ { x , x + 1 } , U_ { x , x + 1 } ] = U_ { x , x + 1 }$, and $[ E_ { x , x + 1 } , U_ { x , x + 1 } ^ \dagger ] = - U_ { x , x + 1 } ^ \dagger$.
In the QLM version of this LGT,
the first commutation relation is modified to $[U _ { x , x + 1 }, U _ { x , x + 1 }^{\dagger}] = 2 E _ { x , x + 1 }$. 
In analogy with quantum angular momentum operators, we can write $U _ {  x , x + 1 } = S _ {  x , x + 1 } ^ { + }$, $U _ {  x , x + 1 } ^ { \dagger } = S _ {  x , x + 1 } ^ { - }$, and $ E _ {  x , x + 1 } = S _ {  x , x + 1 } ^ { 3 }$, so that each link is in a spin-$S$ representation with $S = 0, 1/2, 1, ...$.
The Hamiltonians of the QLM and the LGT look exactly the same, but the link variables and the sizes of the Hilbert spaces are different. If the number of sites on the lattice is finite, then the QLM Hilbert space is finite whereas the LGT Hilbert space is infinite.
The physical Hilbert space of the QLM is defined through the Gauss law $\widetilde { G } _ { x } \ket{\text{phys}} = 0$, with the Gauss-law operator defined in the main text
and $\ket{\text{phys}}$ any state in the physical Hilbert space. 

\section{\label{sec:molecule}Molecular dipole-dipole interaction}

The molecular dipole-dipole interaction~\cite{2015famr.book....3W} is
\begin{equation}
\begin{aligned}
    V 
    =  \frac{1}{2}\sum_{i, j} \sum_{\alpha, \beta, \gamma, \eta} V_{i,j}^{\alpha, \beta; \gamma, \eta} b^\dagger_{i, \gamma} b^\dagger_{j, \eta} b_{j, \beta} b_{i, \alpha}
	=  \frac{1}{2} \sum_{i j} \hat { V } _ { i j }
\end{aligned}
\end{equation}
where
\begin{equation}
\begin{aligned}
     \hat { V } _ { i j } 
	= & \frac { 1 } { 4 \pi \epsilon _ { 0 } } \frac { \hat { \mathbf { d } } _ { i } \cdot \hat { \mathbf { d } } _ { j } - 3 \left( \hat { \mathbf { d } } _ { i } \cdot \hat { r } _ { i j } \right) \left( \hat { \mathbf { d } } _ { j } \cdot \hat { r } _ { i j } \right) } { r _ { i j } ^ { 3 } } \\
	= & \frac { - \sqrt { 6 } } { 4 \pi \epsilon _ { 0 } r _ { i j } ^ { 3 } } \sum _ { p = - 2 } ^ { 2 } ( - 1 ) ^ { p } T _ { - p } ^ { 2 } ( \mathbf { C } ) T _ { p } ^ { 2 } \left( \hat { \mathbf { d } } _ { i } , \hat { \mathbf { d } } _ { j } \right)
\end{aligned}
\end{equation}
is the interaction between two molecules at positions $\mathbf{r}_i$ and $\mathbf{r}_j$, $\mathbf{r}_{i j} \equiv \mathbf{r}_i - \mathbf{r}_j$ is the vector connecting these two molecules, $\hat{r}_{i j} = \mathbf{r}_{i j} / r_{i j}$ is the directional vector, $\mathbf { d } _ { i }$ is the dipole operator of molecule $i$, and $\mathbf { d } _ { j }$ is that of molecule $j$. The functions $T _ { - p } ^ { 2 } ( \mathbf { C } )$ are proportional to spherical harmonics $Y_{2,-p}$
\begin{equation}
	T _ { 0 } ^ { 2 } ( \mathbf { C } ) = \frac{\left( 3 \cos ^ { 2 } \theta_ { i j } - 1 \right)}{2} = 2 \sqrt{\frac{\pi}{5}} Y_{2,0} \left( \theta_ { i j }, \phi_ { i j } \right)
	,
\end{equation}
\begin{equation}
\begin{aligned}
	T _ { \pm 1 } ^ { 2 } ( \mathbf { C } ) = & \mp \sqrt{\frac{3}{2}} e^{ \pm i \phi_{ij}} \sin {\theta_{ij}} \cos \theta_{ij} \\ = & 2 \sqrt{\frac{\pi}{5}} Y_{2,\pm 1} \left( \theta_ { i j }, \phi_ { i j } \right)
	,
\end{aligned}
\end{equation}
\begin{equation}
	T _ { \pm 2 } ^ { 2 } ( \mathbf { C } ) = \sqrt{\frac{3}{8}} e^{ \pm 2 i \phi_{ij}} \sin^2 {\theta_{ij}} = 2 \sqrt{\frac{\pi}{5}} Y_{2,\pm 2} \left( \theta_ { i j }, \phi_ { i j } \right)
	,
\end{equation}
where the polar and azimuthal angles $\theta_{ij}$ and $\phi_{ij}$ are measured with respect to the quantization axis (which we assume to be defined by the direction of a strong uniform magnetic field in experiments with small or zero dc electric field). The quantization axis does not specify a second direction so the azimuthal angles $\phi_{ij}$ are defined only up to an overall offset. $T _ { p } ^ { 2 } \left( \hat { \mathbf { d } } _ { i } , \hat { \mathbf { d } } _ { j } \right)$ are rank-2 tensor operators
\begin{equation}
	T _ { 0 } ^ { 2 } \left( \hat { \mathbf { d } } _ { i } , \hat { \mathbf { d } } _ { j } \right) = \frac { 2 } { \sqrt { 6 } } \left[ \hat { d } _ { 0 } ^ { i } \hat { d } _ { 0 } ^ { j } + \frac { \hat { d } _ { + 1 } ^ { i } \hat { d } _ { - 1 } ^ { j } + \hat { d } _ { - 1 } ^ { i } \hat { d } _ { + 1 } ^ { j } } { 2 } \right] 
	,
\end{equation}
\begin{equation}
	T _ { \pm 1 } ^ { 2 } \left( \hat { \mathbf { d } } _ { i } , \hat { \mathbf { d } } _ { j } \right) = \frac{ \hat { d } _ { 0 } ^ { i } \hat { d } _ { \pm 1 } ^ { j } + \hat { d } _ { \pm 1 } ^ { i } \hat { d } _ { 0 } ^ { j } }{\sqrt{2}}
	,
\end{equation}
\begin{equation}
	T _ { \pm 2 } ^ { 2 } \left( \hat { \mathbf { d } } _ { i } , \hat { \mathbf { d } } _ { j } \right) = \hat { d } _ { \pm 1 } ^ { i } \hat { d } _ { \pm 1 } ^ { j }
	,
\end{equation}
where $\hat{d}_\pm = \hat{d}_x \pm i \hat{d}_y$, $\hat{d}_0 = \hat{d}_z$, $\hat{d}_{\pm 1} = \mp (\hat{d}_x \pm i \hat{d}_y) / \sqrt{2} $, $\hat{d}_{+1} = - \hat{d}_+ / \sqrt{2}$, $\hat{d}_{-1} = \hat{d}_- /\sqrt{2}$. It is worth noting the minus sign in the relation $\hat{d}_{+1}^\dagger = - \hat{d}_{-1}$.

The matrix elements of $\hat{d}_q$ for a given molecule are
\begin{equation}
\begin{aligned}
 & \left\langle N ^ { \prime } , m _ { N } ^ { \prime } \left| \hat { d } _ { q } \right| N , m _ { N } \right\rangle \\
= & d ( - 1 ) ^ { m _ { N }' } \sqrt { \left( 2 N ^ { \prime } + 1 \right) ( 2 N + 1 ) } \\
& \times \left( \begin{array} { c c c } { N' } & { 1 } & { N } \\ { - m _ { N } ' } & { q } & { m _ { N } } \end{array} \right) \left( \begin{array} { c c c } { N ^ { \prime } } & { 1 } & { N } \\ { 0 } & { 0 } & { 0 } \end{array} \right)
,
\end{aligned}
\end{equation}
where the parentheses are Wigner 3-$j$ symbols and $d$ is the electric dipole moment of the molecule. Dipole selection rules $\Delta N = N' - N = \pm 1$ and $\Delta m_N = m_{N}' - m_N = 0, \pm 1$ are required to have a nonzero matrix element explicitly by the second Wigner 3-$j$ symbol.

We only consider $N = 0$ and $N = 1$ states. States with $N \ge 2$ are naturally off-resonant from the initialized configurations. We introduce in the main text the notation for these four states $\ket{a} \equiv \ket{0, 0}$, $\ket{b} \equiv \ket{1, -1}$, $\ket{c} \equiv \ket{1, 0}$ and $\ket{d} \equiv \ket{1, 1}$ for each molecule.
Without external electric fields, magnetic fields, or laser fields, and ignoring internal nuclear structures (i.e., only considering the rotational kinetic energy), states $\ket{b}$, $\ket{c}$, and $\ket{d}$ of a single molecule are degenerate and their energy is greater than the energy of $\ket{a}$ by $2 h B_{\mathrm{rot}}$,
where $B_{\mathrm{rot}}$ is the rotational constant.

Single-molecule nonvanishing dipole matrix elements within the four-state subspace stated above are
\begin{equation}
	\left\langle a \left| \hat { d } _ { +1 } \right| b \right\rangle = \left\langle 0 , 0 \left| \hat { d } _ { +1 } \right| 1 , -1 \right\rangle = -\frac{1}{\sqrt{3}} d
	,
\end{equation}
\begin{equation}
	\left\langle a \left| \hat { d } _ { 0 } \right| c \right\rangle = \left\langle 0 , 0 \left| \hat { d } _ { 0 } \right| 1 , 0 \right\rangle = \frac{1}{\sqrt{3}} d
	,
\end{equation}
\begin{equation}
	\left\langle a \left| \hat { d } _ { -1 } \right| d \right\rangle = \left\langle 0 , 0 \left| \hat { d } _ { -1 } \right| 1 , +1 \right\rangle = -\frac{1}{\sqrt{3}} d
	,
\end{equation}
and their complex conjugates.

Nonzero matrix elements of the dipole-dipole interaction between a pair of molecules in any combination of single-molecule states can be easily calculated by multiplying single-molecule matrix elements.
Nonzero matrix elements are
\begin{equation}
\begin{aligned}
	V_{i, j}^{a, a; b, b} = \paren{ V_{i, j}^{b, b; a, a} }^* = - \frac { 1 } { 4 \pi \epsilon _ { 0 } r _ {i, j} ^ { 3 } } e^{ 2 i \phi_{i, j}} \sin^2 {\theta_{i, j}} \frac{1}{2} d^2
	,
\end{aligned}
\end{equation}
\begin{equation}
\begin{aligned}
	V_{i, j}^{a, a; c, c}  = \paren{ V_{i, j}^{c, c; a, a} }^* = -  \frac { 1 } { 4 \pi \epsilon _ { 0 } r _ {i, j} ^ { 3 } } \left( 3 \cos ^ { 2 } \theta_ {i, j } - 1 \right) \frac{1}{3} d^2
	,
\end{aligned}
\end{equation}
\begin{equation}
\begin{aligned}
	V_{i, j}^{a, a; d, d}  = \paren{ V_{i, j}^{d, d; a, a} }^* =  - \frac { 1 } { 4 \pi \epsilon _ { 0 } r _ {i, j} ^ { 3 } } e^{ - 2 i \phi_{i, j}} \sin^2 {\theta_{i, j}} \frac{1}{2} d^2
	,
\end{aligned}
\end{equation}
\begin{equation}
\begin{aligned}
	& V_{i, j}^{a, a; b, c}  = V_{i, j}^{a, a; c, b} =  \paren{ V_{i, j}^{b, c; a, a} }^*=  \paren{ V_{i, j}^{c, b; a, a} }^* \\ = & - \frac { 1 } { 4 \pi \epsilon _ { 0 } r _ {i, j} ^ { 3 } } e^{  i \phi_{i, j}} \sin {\theta_{i, j}} \cos {\theta_{i, j}} \frac{1}{\sqrt{2}} d^2
	,
\end{aligned}
\end{equation}
\begin{equation}
\begin{aligned}
	& V_{i, j}^{a, a; b, d} = V_{i, j}^{a, a; d, b} =  \paren{ V_{i, j}^{b, d; a, a} }^*=  \paren{ V_{i, j}^{d, b; a, a} }^* \\ = & -  \frac { 1 } { 4 \pi \epsilon _ { 0 } r _ {i, j} ^ { 3 } } \frac{\left( 3 \cos ^ { 2 } \theta_ {i, j } - 1 \right)}{2} \frac{1}{3} d^2
	,
\end{aligned}
\end{equation}
\begin{equation}
\begin{aligned}
	& V_{i, j}^{a, a; c, d} = V_{i, j}^{a, a; d, c} =  \paren{ V_{i, j}^{c, d; a, a} }^*=  \paren{ V_{i, j}^{d, c; a, a} }^* \\ = & \frac { 1 } { 4 \pi \epsilon _ { 0 } r _ {i, j} ^ { 3 } } e^{ - i \phi_{i, j}} \sin {\theta_{i, j}} \cos {\theta_{i, j}} \frac{1}{\sqrt{2}} d^2
	,
\end{aligned}
\end{equation}
\begin{equation}
\begin{aligned}
	V_{i, j}^{a, b; b, a} = V_{i, j}^{b, a; a, b} = \frac { 1 } { 4 \pi \epsilon _ { 0 } r _ {i, j} ^ { 3 } } \frac{\left( 3 \cos ^ { 2 } \theta_ {i, j } - 1 \right)}{2} \frac{1}{3} d^2
	,
\end{aligned}
\end{equation}
\begin{equation}
\begin{aligned}
	V_{i, j}^{a, c; c, a} = V_{i, j}^{c, a; a, c} = - \frac { 1 } { 4 \pi \epsilon _ { 0 } r _ {i, j} ^ { 3 } } \left( 3 \cos ^ { 2 } \theta_ {i, j } - 1 \right) \frac{1}{3} d^2
	,
\end{aligned}
\end{equation}
\begin{equation}
\begin{aligned}
	V_{i, j}^{a, d; d, a} = V_{i, j}^{d, a; a, d} = \frac { 1 } { 4 \pi \epsilon _ { 0 } r _ {i, j} ^ { 3 } } \frac{\left( 3 \cos ^ { 2 } \theta_ {i, j } - 1 \right)}{2} \frac{1}{3} d^2
	,
\end{aligned}
\end{equation}
\begin{equation}
\begin{aligned}
	& V_{i, j}^{a, b; c, a} = V_{i, j}^{b, a; a, c} = \paren{V_{i, j}^{c, a; a, b}}^* = \paren{V_{i, j}^{a, c; b, a}}^* \\ = & - \frac { 1 } { 4 \pi \epsilon _ { 0 } r _ {i, j} ^ { 3 } } e^{ - i \phi_{i, j}} \sin {\theta_{i, j}} \cos {\theta_{i, j}} \frac{1}{\sqrt{2}} d^2
	,
\end{aligned}
\end{equation}
\begin{equation}
\begin{aligned}
	& V_{i, j}^{a, b; d, a} = V_{i, j}^{b, a; a, d} = \paren{V_{i, j}^{d, a; a, b}}^* = \paren{V_{i, j}^{a, d; b, a}}^* \\ = & \frac { 1 } { 4 \pi \epsilon _ { 0 } r _ {i, j} ^ { 3 } } e^{ - 2 i \phi_{i, j}} \sin^2 {\theta_{i, j}} \frac{1}{2} d^2
	,
\end{aligned}
\end{equation}
\begin{equation}
\begin{aligned}
	& V_{i, j}^{a, c; d, a} = V_{i, j}^{c, a; a, d} = \paren{V_{i, j}^{d, a; a, c}}^* = \paren{V_{i, j}^{a, d; c, a}}^* \\ = & \frac { 1 } { 4 \pi \epsilon _ { 0 } r _ {i, j} ^ { 3 } } e^{ - i \phi_{i, j}} \sin {\theta_{i, j}} \cos {\theta_{i, j}} \frac{1}{\sqrt{2}} d^2
	.
\end{aligned}
\end{equation}

\section{\label{sec:molecule2}Molecular internal state energies and their experimental control}

The dipole-dipole interactions described in the previous section provide the fundamental mechanism by which dynamics can proceed and by which the densities of various internal states can evolve in the considered system of molecular ``spins'' fixed in place. In the mapping to the QLM, this will provide a mechanism for ``fermions'' (or ``charges'') hopping, as represented by spin excitations (i.e., hard-core bosons) being exchanged between different fixed molecules.

Equally important to our proposed framework is the ability to restrict these dipolar exchange processes in a controlled way. Specifically, by imposing energetic constraints on the various internal-state configurations of the molecules, we can effectively impose gauge invariance or enforce Gauss's law, by only allowing processes that correlate the hopping of fermions between ``sites'' with the modification of the spin that resides on the intervening ``link.'' In this approach, such energetic constraints are imposed directly on the molecules through the single-particle terms ($H_0$) of the DMH. These energy terms $\epsilon_{i,\alpha}$ of the DMH depend in general on both the molecule position (labeled by the index $i$) and the internal rotational level (denoted by $\alpha$). First off, for typical experiments on ultracold molecules operating at large magnetic fields (near the field values used for magnetoassociation of the atomic constituents), rotational-level-dependent energy terms arise due to the weak coupling between molecular rotation and the hyperfine (nuclear) degrees of freedom~\cite{Gorshkov-DipolarMagnetismPRA-2011}. These naturally arising shifts to the various rotational levels serve to break the degeneracy of the $N=1$ rotational manifold at the scale of $\sim 10-100$~kHz, even in zero electric field. In addition to this, the energies of the rotational levels can be modified globally through the addition of weak dc electric fields or off-resonant (and polarized) microwave fields. These can be used, e.g., for the purpose of shifting particular rotational sublevels of the $N=1$ manifold by a large amount so as to decouple it from near-resonant dipole-driven dynamics.

Finally, and most central to the proposed approach, spatially resolved control of the internal-state energies can be engineered by direct optical addressing, using level-dependent ac Stark shifts to tune the internal state energies. Such an ability to locally address individual molecules arises naturally in implementations based on microtrapped arrays~\cite{Anderegg1156}, but could also be achieved by projecting tailored laser patterns onto lattice-trapped samples.

In the proposed scheme, the positions of all molecules are fixed, and the total number of molecules as well as the total number of molecules in the $N=0$ rotational ground state, or level $\ket{a}$, are conserved. As such, the full tuning of all relevant configurations of molecules can be accomplished through local and level-dependent control of the differential (with respect to $\ket{a}$) ac Stark shift of the utilized $N=1$ sublevels. In general, molecules play host to a strongly anisotropic and rotational level-dependent ac polarizability~\cite{Neyenhuis-Polarizability-12}. By control of the local laser intensity and polarization (with respect to the quantization axis, here assumed to be defined by a quantization magnetic field), a large differential tuning of these energies is available for almost any laser wavelength.

For a complete and general control, we consider addressing the array of molecules with a control laser that is tuned near a narrow optical transition from the molecular ground state to a relatively long-lived electronic excited state. For commonly used bialkalis such as NaRb or KRb, this could for example relate to transitions of the form $\ket{X^1 \Sigma , \nu = 0, N = 1, m_N} \rightarrow \ket{b^3 \Pi_{0^+}, \nu = 0, N = 0, m_N = 0}$, characterized by kilohertz-level linewidths~\cite{Kobayashi,Bloch-mol-tuneout,NormHopf-2019}. In particular, for gigahertz-scale detunings from such a transition, local control of laser intensity and polarization would provide complete control over all relevant differential rotational level-dependent energies $\epsilon_{i,\alpha}$ of the DMH, owing to dipole selection rules. Shifts at the necessary scales (even up to megahertz order) can be accommodated with modest optical powers in scenarios based on local projection of tightly focused lasers.
For alternative realizations based on arrays of Rydberg atoms, we note that the control of internal state-dependent energies via local state-dependent ac Stark shifts has already been demonstrated~\cite{Browaeys-opticalshift}.

The local detection of the various molecular internal states could be accomplished, e.g., by mapping them onto different atomic levels (in a reversal of the stimulated Raman adiabatic passage process) followed by imaging of the atoms, or alternatively by extensions of direct molecular detection methods~\cite{WangImage}. Similar capabilities will be equally critical to the development of molecules as qubits or qudit architectures for applications in quantum information science.

\section{Quasidegenerate effective Hamiltonians}

Before the construction of QLM Hamiltonians, we first introduce the method of quasidegenerate effective Hamiltonians~\cite{winkler2003spin}. This method is a perturbative way of calculating an effective Hamiltonian that will yield similar dynamics as the original Hamiltonian $H = H_0 + V$ in a subspace $\alpha$ which we are interested in. Eigenstates $\ket{m, \alpha} \in \alpha$ of $H_0$ are given by $H_0 \ket{m, \alpha} = E_{m \alpha} \ket{m, \alpha}$. $E_{m \alpha}$ for different $m$'s are nearly degenerate and small variations are allowed. Eigenvalues of eigenstates of $H_0$ outside the subspace $\alpha$ are separated from $E_{m \alpha}$. To second order, the matrix elements of the effective Hamiltonian for $\alpha$ are
\begin{equation}
\begin{aligned}
& \left\langle m, \alpha \left| H _ { \mathrm { eff } } ^ { \alpha } \right| n, \alpha \right\rangle \\ =  & \ E _ { m \alpha } \delta _ { m, n } + \langle m , \alpha | V | n , \alpha \rangle \\ & + \frac { 1 } { 2 } \sum _ { l , \gamma \neq \alpha } \langle m , \alpha | V | l , \gamma \rangle \langle l , \gamma | V | n , \alpha \rangle \\ & \times \left[ \frac { 1 } { E _ { m \alpha } - E _ { l \gamma } } + \frac { 1 } { E _ { n \alpha } - E _ { l \gamma } } \right] + \cdots
	,
\end{aligned}
\end{equation}
where greek letters label subspaces and roman letters label states.

In our case of realizing QLMs, $\alpha$ is chosen as the subspace of DMHs that maps to the physical Hilbert space of QLMs. There are no two-body interactions in the QLM Hamiltonian so there should be no such terms in the effective Hamiltonian, either. We describe here how to suppress these terms in the effective Hamiltonian.
The dipole-dipole interaction decays as a power law of $r^{-3}$ but it exists even for two molecules far away from each other.
We choose parameters in Table \ref{table:energy_conditions_S=1/2_01}, as explained below, such that only our wanted DMH states are nearly degenerate and $\langle i , \alpha | V | j , \alpha \rangle$ always vanishes.
In addition, for the two molecules extremely far away from each other such that their dipole-dipole interaction is much smaller than $w$, $m$, and $g^2$, then their dipole-dipole interaction can be neglected and the requirement of energy separations can be loosened in this case.

\section{Construction of Fermions from hard-core bosons}

We map the fermion operators in QLMs to spin operators through a Jordan-Wigner transformation, which will be further related to the hard core boson operators in the DMH.

According to the mapping between fermion site states and dipolar molecule states in Table I in the main text for both spin 1/2 and spin 1, the occupied fermion site is always mapped to $|a\rangle$ while the unoccupied fermion site is mapped to either $|b\rangle$ or $|c\rangle$. In our setup, each ``fermion'' site in the DMH is either occupied or unoccupied. Namely, $b^\dagger_{x , a} b_{x , a} + b^\dagger_{x , \downarrow} b_{x , \downarrow} = 1$, where $\downarrow = b$ or $c$. In this situation, each molecule is a two-level system and analogous to spin-$1/2$.

Since the internal state excitations of dipolar molecules can be described as hard core bosons, in one spatial dimension a Jordan-Wigner transformation can map the hardcore boson states to the fermion states while preserving the locality of local operators. The Jordan-Wigner transformation takes fermion operators to spin-$1/2$ operators, $\psi^\dagger_x \prod_{\beta = 1}^{x - 1} e ^ { i \pi \psi _ { \beta } ^ { \dagger } \psi _ { \beta } } = S^+_x$, $\psi_x \prod_{\beta = 1}^{x - 1} e ^ { - i \pi \psi _ { \beta } ^ { \dagger } \psi _ { \beta } } = S^-_x$. As a result, $\psi^\dagger_x \psi_x = S^3_x + 1/2$ and $\psi^\dagger_{x} \psi_{x+1} = S_x^+ S_{x+1}^-$. The spin-$1/2$ operators mentioned above, $S_{x}^+$, $S_{x}^-$, and $S_{x}^3$, have no relation with the spin operators on quantum links. Similarly, they are not to be confused with the actual site identifiers of the form $S_x$ and $S_{x+1}$ as introduced in the main text.

Finally, we have relations between spin operators and hard core bosonic operators in the DMH, which are $S^+_x  =  b^\dagger_{x , a} b_{x , \downarrow}$, $S^-_x =  b^\dagger_{x , \downarrow} b_{x , a}$, and $S^3_{x} = ( b^\dagger_{x, a} b_{x, a} - b^\dagger_{x, \downarrow} b_{x, \downarrow} ) / 2$. They give rise to $\psi^\dagger_x \psi_x = S^3_x + 1/2 =  ( b^\dagger_{x , a} b_{x , a} - b^\dagger_{x , \downarrow} b^\dagger_{x , \downarrow} + 1) / 2$ and $\psi^\dagger_{x} \psi_{x+1} = b^\dagger_{x , a} b_{x , \downarrow} b^\dagger_{x+1 , \downarrow} b_{x+1 , a}$.

\section{Construction of the \texorpdfstring{$S = 1/2$}{} QLM Hamlitonian}

For the $S = 1/2$ QLM, the link operators can be mapped to hard core bosonic operators as follows: $U _ {  x , x + 1 } = S _ {  x , x + 1 } ^ { + } = b_{(x, x+1), d}^\dagger b_{(x, x+1), b}$, $U _ {  x , x + 1 } ^ { \dagger } = S _ {  x , x + 1 } ^ { - } = b_{(x, x+1) , b}^\dagger b_{(x, x+1), d}$ and $E _ {  x , x + 1 } = S _ {  x , x + 1 } ^ { 3 } = (b_{(x, x+1), d}^\dagger b_{(x, x+1), d} - b_{(x, x+1), b}^\dagger b_{(x, x+1), b}) / 2$. With the mappings given above, the $S = 1/2$ QLM Hamiltonian can be written in the dipolar molecular operators as
\begin{equation}
\begin{aligned}
     & H \\
    = & - w \sum _ { x } \left[ b^\dagger_{x , a} b_{x , b} b_{(x, x+1) , d}^\dagger b_{(x, x+1) , b} b _ { x + 1, b } ^ \dagger b _ { x + 1, a } + \text{H.c.} \right] \\
    & + m \sum _ { x } ( - 1 ) ^ { x } \frac{1}{2} \paren{ b^\dagger_{x , a} b_{x , a} - b^\dagger_{x , b} b_{x , b} + 1} + \frac { g ^ { 2 } } { 2 } \sum _ { x } \frac{1}{4}
    ,
\end{aligned}
\end{equation}
where the last term is a constant and can be discarded.

\begin{figure}[b]
  \centering
  \includegraphics[width=0.8\linewidth]{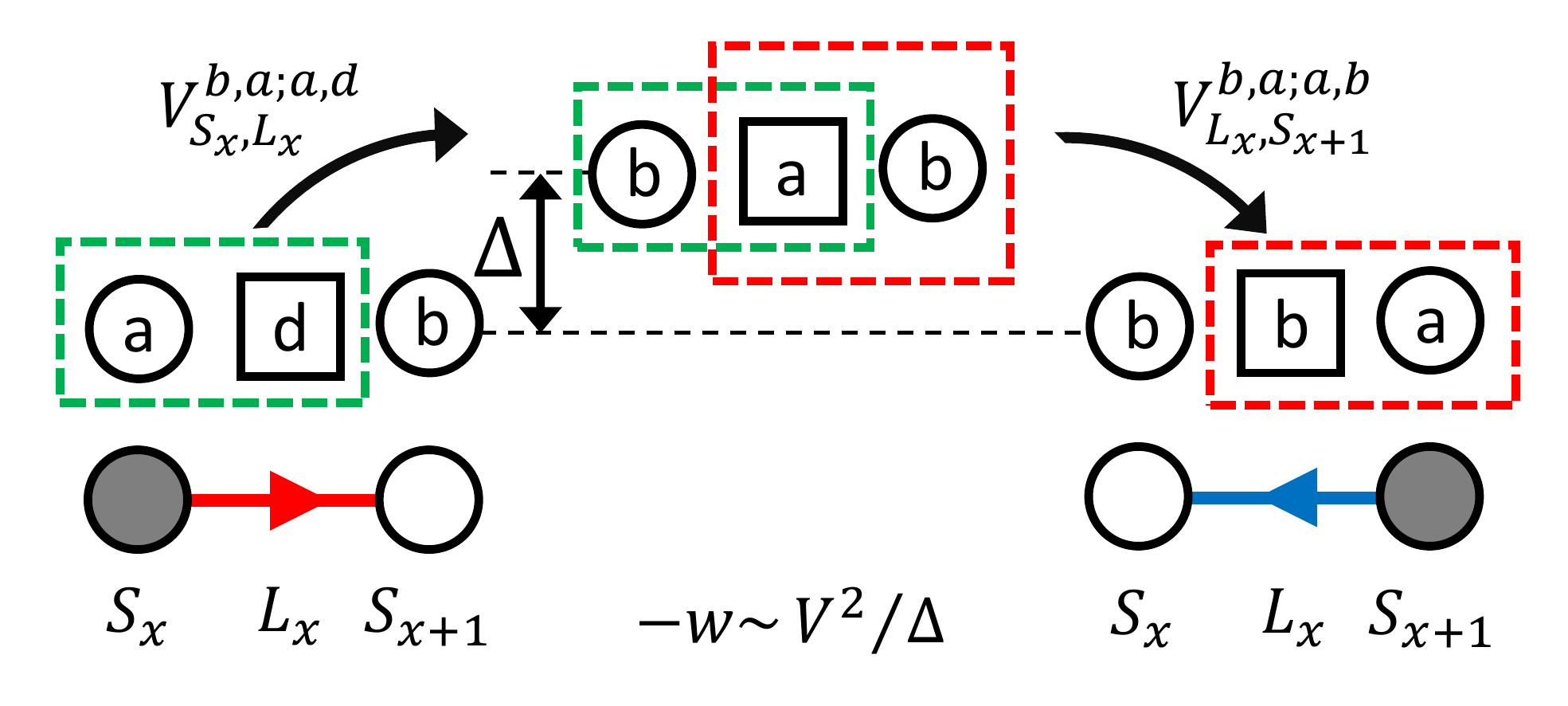}
  \caption{Schematic hopping on $S_x$, $L_x$, and $S_{x+1}$ through a virtual intermediate molecular configuration. The QLM states and the molecular levels are shown together for the initial and final configurations, respectively. The key hopping process of the QLM is achieved through a second-order dipolar exchange process, where the correlation between matter hopping and the changes to the link spin (Gauss's law) is imposed through energetic constraints.
  }
  \label{fig:hopping}
\end{figure}

The hopping process $\ket{a}_{S_{x}}\ket{d}_{L_x}\ket{b}_{S_{x+1}} \xrightarrow[]{\text{virtual}}\ket{b}_{S_x}\ket{a}_{L_x}\ket{b}_{S_{x+1}} \xrightarrow[]{\text{virtual}} \ket{b}_{S_x}\ket{b}_{L_x}\ket{a}_{S_{x+1}}$ described in the main text (where $\ket{b}_{S_x}\ket{a}_{L_x}\ket{b}_{S_{x+1}}$ is a virtual intermediary) is illustrated schematically in Fig.~\ref{fig:hopping}.

We choose $\epsilon_{i, \alpha}$ introduced in Eq.~2 in the main text as is shown in Table~\ref{table:energy_conditions_S=1/2_01} for the $n$-th unit cell. $\Delta_{1, n}$, $\Delta_{2, n}$, $\delta_{1, n}$, and $\delta_{2, n}$ for any $n$ are at the order of $\Delta$ mentioned in the main text and are specified in numerical simulations.

For a molecule in the $N = 0$ ($| a \rangle$) state and another molecule in an $N = 1$ state, there are self-interactions to second order, a virtual process in which a state first hops to an intermediate state and then hops back to the initial state. This type of self-interaction always exists in principle for such pairs of molecules, no matter how far they are from each other in space, although the self-interaction decays as $r^{-6}$. If the distance between the two molecules is far away enough such that the self-interaction is much smaller than relevant energy scales in the QLM, $w$, $m$, and $g^2$, then those self-interactions can be neglected. For the $S = 1/2$ QLM, we use a molecular chain with roughly uniform spacing between adjacent molecules (within the order of magnitude), which are specified later in the numerical details. The second-order self-interaction between next-nearest molecules (such as $S_1$ and $S_2$) is about $w / 2^6 = w / 64 \ll w$ because of the $r^{-6}$ scaling. Therefore, we can safely neglect self-interactions between next-nearest molecules and those between two molecules that are even farther apart. We only consider self-interactions between the nearest molecules (such as $S_1$ and $L_1$) which are at the order of $w$.

\onecolumngrid
\begin{center}
\begin{table}[t]
\begin{center}
\begingroup
\setlength{\tabcolsep}{6pt}
\renewcommand{\arraystretch}{2}
\begin{tabular}{c c}
	\hline\hline
	Molecule State Energy & Value \\
	\hline
	$\epsilon_{S_{2 n + 1}, a}$ &
	$- m - \Sigma_{S_{2n+1}, a; L_{2n+1}, b} - \Sigma_{L_{2n+2}, b; S_{2n+3}, a}$
	\\
	$\epsilon_{S_{2 n + 1}, b}$ & $2 h B + \delta_{1, n}$ \\
	\hline
	$\epsilon_{L_{2 n + 1}, a}$ & $0$ \\
	$\epsilon_{L_{2 n + 1}, 1/2}$ ($\epsilon_{L_{2 n + 1}, d}$) &\begingroup \renewcommand{\arraystretch}{1} \begin{tabular}{@{}c@{}} $2 h B + \Delta_{1, n} + g^2 / 2$ \\  $- \Sigma_{S_{2n+1}, a; L_{2n+1}, d} - \Sigma_{L_{2n+2}, d; S_{2n+3}, a} + \Sigma_{S_{2n+1}, a; L_{2n+1}, b} + \Sigma_{L_{2n+2}, b; S_{2n+3}, a} $ \end{tabular} \endgroup \\
	$\epsilon_{L_{2 n + 1}, -1/2}$ ($\epsilon_{L_{2 n + 1}, b}$) & $2 h B + \Delta_{1, n} + \paren{ \delta_{2, n} - \delta_{1, n} } + g^2 / 2$ \\
	 \hline
	$\epsilon_{S_{2 n + 2}, a}$ & $m - \Sigma_{L_{2n+1}, b; S_{2n+2}, a} - \Sigma_{S_{2n+2}, a; L_{2n+2}, d} $ \\
	$\epsilon_{S_{2 n + 2}, b}$ & $2 h B + \delta_{2, n}$ \\
	\hline
	$\epsilon_{L_{2 n + 2}, a}$ & $0$ \\
	$\epsilon_{L_{2 n + 2}, 1/2}$ ($\epsilon_{L_{2 n + 2}, d}$) & $2 h B + \Delta_{2, n} + g^2 / 2$
	 \\
	 $\epsilon_{L_{2 n + 2}, -1/2}$ ($\epsilon_{L_{2 n + 2}, b}$) & $2 h B + \Delta_{2, n} + \paren{ \delta_{1, n + 1} - \delta_{2, n} } + g^2 / 2$ \\
	 \hline\hline
\end{tabular}
\endgroup
\end{center}
\caption{Energy conditions for the $n$-th unit cell in the $S = 1/2$ QLM, $n = 0, 1, 2,....$ The other molecule states not listed are made off resonant. $B$ is the molecule's rotational constant and $h$ is Planck's constant.}
\label{table:energy_conditions_S=1/2_01}
\end{table}
\end{center}

\twocolumngrid
Self-interactions between molecule $i$ with internal state $\alpha$ and molecule $j$ with internal state $\beta$ can be denoted as
\begin{equation}
    \Sigma_{i, \alpha; j, \beta} \hat{n}_{i, \alpha} \hat{n}_{j, \beta}
    ,
\end{equation}
where
\begin{equation}
\begin{aligned}
    & \Sigma_{i, \alpha; j, \beta} \\
    = & \sum_{\substack{\gamma \ne \alpha \\ \eta \ne \beta} } V_{i, j}^{\alpha, \beta ; \gamma, \eta} V_{i, j}^{\gamma, \eta ; \alpha, \beta} \frac{1}{\epsilon_{i, \alpha} + \epsilon_{j, \beta} - \epsilon_{i, \gamma} - \epsilon_{j, \eta}} \\
    = & \sum_{\substack{\gamma \ne \alpha \\ \eta \ne \beta} } \left| V_{i, j}^{\alpha, \beta ; \gamma, \eta} \right|^2 \frac{1}{\epsilon_{i, \alpha} + \epsilon_{j, \beta} - \epsilon_{i, \gamma} - \epsilon_{j, \eta}}
\end{aligned}
\label{eq:self-interaction_general_value}
\end{equation}
is the coefficient.
From the dipole selection rule, $\Sigma_{i, \alpha; j, \beta} = 0$ if both $\alpha$ and $\beta$ are from $N = 1$, or both are from $N = 0$.
For our purpose, $\Sigma_{i, \alpha; j, \beta}$ only needs to be precise at the order of $O (V^2 / \Delta)$ and any higher orders can be neglected. Therefore, in calculating the denominator in Eq.~(\ref{eq:self-interaction_general_value}), $\epsilon_{i, \alpha} + \epsilon_{j, \beta} - \epsilon_{i, \gamma} - \epsilon_{j, \eta}$, corrections at the order of $m, g^2,\Sigma \ll \Delta$ can be neglected, since these corrections when propagated to Eq.~(\ref{eq:self-interaction_general_value}) are at most at the order of $O (V^3 / \Delta^2)$. Note that we are interested in regions of parameter space where $m$, $g^2$, and $w$ are comparable, and also, $\Sigma \sim O (w)$.

We explain in detail how to suppress the nearest self-interaction for $S = 1/2$. The nearest self-interactions for $S = 1/2$ are products of two number operators, as is shown in Eq.~(\ref{eq:self-interaction_general_value}). These products of number operators, in the physical Hilbert space, can be rewritten into one-body terms which are just single number operators using the Gauss' law. Single number operators are one-body potentials and we can compensate them by adding laser light potentials of the opposite values.

The Gauss law on the site $S_{2n + 2}$, written in terms of molecule number operators, is
\begin{equation}
	\hat{n}_{S_{2n + 2}, a} - \hat{n}_{L_{2n + 2}, d} + \hat{n}_{L_{2n + 1}, d} = 0
	,
\end{equation}
where we have used constraints on each molecular position:
\begin{equation}
	\hat{n}_{S_{2n+2}, a} + \hat{n}_{S_{2n+2}, b} = 1
	,
\end{equation}
\begin{equation}
	\hat{n}_{L_{2n+1}, b} + \hat{n}_{L_{2n+1}, d} = 1
	,
\end{equation}
\begin{equation}
	\hat{n}_{L_{2n+2}, b} + \hat{n}_{L_{2n+2}, d} = 1
	.
\end{equation}
We can simplify the two-body interaction terms to one-body potentials by using these constraints on number operators. From the Gauss law,
\begin{equation}
	\hat{n}_{S_{2n+2}, a} - \hat{n}_{L_{2n+2}, d} = - \hat{n}_{L_{2n+1}, d}
	,
\end{equation}
we square it:
\begin{equation}
	\hat{n}_{S_{2n+2}, a}^2 + \hat{n}_{L_{2n+2}, d}^2 - 2 \hat{n}_{S_{2n+2}, a} \hat{n}_{L_{2n+2}, d} = \hat{n}_{L_{2n+1}, d}^2
	.
\end{equation}
Because $\hat{n}_{i, \alpha}^2 = \hat{n}_{i, \alpha}$ for hard-core bosonic states,
\begin{equation}
	\hat{n}_{S_{2n+2}, a} + \hat{n}_{L_{2n+2}, d} - 2 \hat{n}_{S_{2n+2}, a} \hat{n}_{L_{2n+2}, d} = \hat{n}_{L_{2n+1}, d}
	,
\end{equation}
and we arrive at
\begin{equation}
	\hat{n}_{S_{2n+2}, a} \hat{n}_{L_{2n+2}, d} = \hat{n}_{S_{2n+2}, a}
	.
\label{eq:gauss_s2n+2_1}
\end{equation}
Similarly,
\begin{equation}
	\hat{n}_{S_{2n+2}, a} \hat{n}_{L_{2n+1}, d} = 0
	,
\label{eq:gauss_s2n+2_2}
\end{equation}
\begin{equation}
	\hat{n}_{S_{2n+2}, a} \hat{n}_{L_{2n+2}, b} = 0
	,
\label{eq:gauss_s2n+2_3}
\end{equation}
\begin{equation}
	\hat{n}_{S_{2n+2}, a} \hat{n}_{L_{2n+1}, b} = 	\hat{n}_{S_{2n+2}, a}
	.
\label{eq:gauss_s2n+2_4}
\end{equation}

The Gauss law on the site $S_{2n+1}$, written in terms of molecule number operators, is
\begin{equation}
	\hat{n}_{S_{2n+1}, a} - \hat{n}_{L_{2n+1}, d} + \hat{n}_{L_{2n}, d} = 1
	,
\end{equation}
where we have used constraints on each molecular position
\begin{equation}
	\hat{n}_{S_{2n+1}, a} + \hat{n}_{S_{2n+1}, b} = 1
	,
\end{equation}
\begin{equation}
	\hat{n}_{L_{2n+1}, b} + \hat{n}_{L_{2n+1}, d} = 1
	,
\end{equation}
\begin{equation}
	\hat{n}_{L_{2n}, b} + \hat{n}_{L_{2n}, d} = 1
	.
\end{equation}
We can derive that
\begin{equation}
	\hat{n}_{S_{2n+1}, a} \hat{n}_{L_{2n+1}, d} = \hat{n}_{L_{2n+1}, d}
	,
\label{eq:gauss_s2n+1_1}
\end{equation}
\begin{equation}
	\hat{n}_{S_{2n+1}, a} \hat{n}_{L_{2n}, d} = \hat{n}_{L_{2n}, d}
	,
\label{eq:gauss_s2n+1_2}
\end{equation}
\begin{equation}
	\hat{n}_{S_{2n+1}, a} \hat{n}_{L_{2n+1}, b} = \hat{n}_{S_{2n+1}, a} - \hat{n}_{L_{2n+1}, d}
	,
\label{eq:gauss_s2n+1_3}
\end{equation}
\begin{equation}
	\hat{n}_{S_{2n+1}, a} \hat{n}_{L_{2n}, b} = \hat{n}_{S_{2n+1}, a} - \hat{n}_{L_{2n}, d}
	.
\label{eq:gauss_s2n+1_4}
\end{equation}
The left-hand sides of Eqs.~(\ref{eq:gauss_s2n+2_1})-(\ref{eq:gauss_s2n+2_4}) and Eqs.~(\ref{eq:gauss_s2n+1_1})-(\ref{eq:gauss_s2n+1_4})
are the only possible combinations of two number operators with nonzero second-order self-interactions. The right-hand sides of Eqs.~(\ref{eq:gauss_s2n+2_1})-(\ref{eq:gauss_s2n+2_4}) and Eqs.~(\ref{eq:gauss_s2n+1_1})-(\ref{eq:gauss_s2n+1_4})
show that the effects of self-interactions are equivalent to one-body terms. These one-body terms from self-interactions are not in the original QLM Hamiltonian, so we want to compensate for them by using laser lights to introduce one-body terms with opposite values. 

To know what values should be used to compensate for the self-interactions, we need to calculate $\Sigma_{i, \alpha; j, \beta}$.
With the general form in Eq.~(\ref{eq:self-interaction_general_value}), we can plug values of $V_{i, j}^{\alpha, \beta ; \gamma, \eta}$ and $\epsilon_{i, \alpha}$ to obtain all of the nearest self-interaction coefficients as follows:
\begin{equation}
\begin{aligned}
	& \Sigma_{L_{2n+1}, b; S_{2n+2}, a} \\
	= & V_{L_{2n+1}, S_{2n+2}}^{b, a; a, b} V_{L_{2n+1}, S_{2n+2}}^{a, b; b, a} \frac{1}{\Delta_{1, n} - \delta_{1, n}} \\
	= & \bracket{ \frac { 1 } { 4 \pi \epsilon _ { 0 } r _ {L_{2n+1}, S_{2n+2}} ^ { 3 } } \frac{\left( 3 \cos ^ { 2 } \theta_ {L_{2n+1}, S_{2n+2}} - 1 \right)}{2} \frac{1}{3} d^2 } ^ 2 \\ & \times \frac{1}{\Delta_{1, n} - \delta_{1, n}}
	,
\end{aligned}
\end{equation}
\begin{equation}
\begin{aligned}
	& \Sigma_{S_{2n+2}, a; L_{2n+2}, d} \\
	= & V_{S_{2n+2}, L_{2n+2}}^{a, d; b, a} V_{S_{2n+2}, L_{2n+2}}^{b, a; a, d} \frac{1}{\Delta_{2,n} - \delta_{2, n}} \\ 
	= & \bracket{ \frac { 1 } { 4 \pi \epsilon _ { 0 } r _ {S_{2n+2}, L_{2n+2}} ^ { 3 } } \sin^2 {\theta_{S_{2n+2}, L_{2n+2}}} \frac{1}{2} d^2 } ^2 \\ & \times \frac{1}{\Delta_{2,n} - \delta_{2, n}}
	,
\end{aligned}
\end{equation}
\begin{equation}
\begin{aligned}
	& \Sigma_{S_{2n+1}, a; L_{2n+1}, d} \\
	= & V_{S_{2n+1}, L_{2n+1}}^{a, d; b, a} V_{S_{2n+1}, L_{2n+1}}^{b, a; a, d} \frac{1}{\Delta_{1, n} - \delta_{1, n}} \\ 
	= & \bracket{ \frac { 1 } { 4 \pi \epsilon _ { 0 } r _ {S_{2n+1}, L_{2n+1}} ^ { 3 } } \sin^2 {\theta_{S_{2n+1}, L_{2n+1}}} \frac{1}{2} d^2 } ^2 \\ & \times \frac{1}{\Delta_{1, n} - \delta_{1, n}}
	,
\end{aligned}
\end{equation}
\begin{equation}
\begin{aligned}
	& \Sigma_{L_{2n+2}, d; S_{2n+3}, a} \\
	= & V_{L_{2n+2}, S_{2n+3}}^{d, a; a, b} V_{L_{2n+2}, S_{2n+3}}^{a, b; d, a} \frac{1}{\Delta_{2, n} - \delta_{1, n+1}} \\ 
	= & \bracket{ \frac { 1 } { 4 \pi \epsilon _ { 0 } r _ {L_{2n+2}, S_{2n+3}} ^ { 3 } } \sin^2 {\theta_{L_{2n+2}, S_{2n+3}}} \frac{1}{2} d^2 } ^2 \\ & \times \frac{1}{\Delta_{2, n} - \delta_{1, n+1}}
	,
\end{aligned}
\end{equation}
\begin{equation}
\begin{aligned}
	& \Sigma_{S_{2n+1}, a; L_{2n+1}, b} \\
	= & V_{S_{2n+1}, L_{2n+1}}^{a, b; b, a} V_{S_{2n+1}, L_{2n+1}}^{b, a; a, b} \frac{1}{\Delta_{1, n} + \delta_{2, n} - 2 \delta_{1, n}} \\ 
	= & \bracket{ \frac { 1 } { 4 \pi \epsilon _ { 0 } r _ { S_{2n+1}, L_{2n+1} } ^ { 3 } } \frac{\left( 3 \cos ^ { 2 } \theta_ {S_{2n+1}, L_{2n+1} } - 1 \right)}{2} \frac{1}{3} d^2 } ^ 2 \\ & \times \frac{1}{\Delta_{1, n} + \delta_{2, n} - 2 \delta_{1, n}}
	,
\end{aligned}
\end{equation}
\begin{equation}
\begin{aligned}
	& \Sigma_{L_{2n+2}, b; S_{2n+3}, a} \\
	= & V_{L_{2n+2}, S_{2n+3}}^{b, a; a, b} V_{L_{2n+2}, S_{2n+3}}^{a, b; b, a} \frac{1}{\Delta_{2, n} - \delta_{2, n}} \\ 
	= & \bracket{ \frac { 1 } { 4 \pi \epsilon _ { 0 } r _ { L_{2n+2}, S_{2n+3} } ^ { 3 } } \frac{\left( 3 \cos ^ { 2 } \theta_ {L_{2n+2}, S_{2n+3} } - 1 \right)}{2} \frac{1}{3} d^2 } ^ 2 \\ & \times \frac{1}{\Delta_{2, n} - \delta_{2, n}}
	.
\end{aligned}
\end{equation}

\section{Construction of the \texorpdfstring{$S = 1$}{} QLM Hamlitonian}
\newpage
The $S = 1$ quantum link model is
\begin{widetext}
\begin{equation}
\begin{aligned}
 H 
= & - w \sum _ { x } \left[ b_{x, a} ^ { \dagger } b_{x, c} S _ {  x , x + 1 } ^ { + } b_{x + 1, c} ^ { \dagger } b_{ x + 1, a}  + b_{x + 1, a} ^ { \dagger } b_{ x + 1, c} S _ {  x , x + 1 } ^ { - } b_{x, c} ^ { \dagger } b_{ x , a} \right] \\ & + m \sum _ { x } ( - 1 ) ^ { x } \frac{1}{2} \paren{ b^\dagger_{x , a} b_{x , a} - b^\dagger_{x , c} b^\dagger_{x , c} + 1} + \frac { g ^ { 2 } } { 2 } \sum _ { x } \paren{ S _ {  x , x + 1 } ^ { 3 } } ^ { 2 } \\
= & - w \sum _ { x } \sqrt{2} \left[ b_{x, a} ^ { \dagger } b_{ x, c}  \paren{ b_{(x, x+1), c}^\dagger b_{(x,x+1), b} + b_{(x, x+1), b}^\dagger b_{(x,x+1), d} } b_{ x + 1, c} ^ { \dagger } b_{ x + 1, a} \right. \\ 
& + \left.b_{ x + 1, a} ^ { \dagger } b_{ x + 1, c} \paren{ b_{(x,x+1), b}^\dagger b_{(x,x+1), c} + b_{(x,x+1), d}^\dagger b_{(x,x+1), b} } b_{ x, c} ^ { \dagger } b_{ x , a} \right] \\ 
& + m \sum _ { x } ( - 1 ) ^ { x } \frac{1}{2} \paren{ b^\dagger_{x , a} b_{x , a} - b^\dagger_{x , c} b^\dagger_{x , c} + 1} + \frac { g ^ { 2 } } { 2 } \sum _ { x } \paren{b_{(x,x+1) , c}^\dagger b_{(x,x+1), c} - b_{(x,x+1), d}^\dagger b_{(x,x+1), d}} ^ { 2}
\label{eq:eff_ham_S=1_01}
\end{aligned}
\end{equation}
\end{widetext}
where the $S = 1$ spin operators are
\begin{equation}
	S _ {  x , x + 1 } ^ { + } = \sqrt{2} \paren{ b_{(x,x+1) , c}^\dagger b_{(x,x+1), b} + b_{(x,x
	+1), b}^\dagger b_{(x,x+1), d} }
	,
\end{equation}
\begin{equation}
	S _ {  x , x + 1 } ^ { - } = \sqrt{2} \paren{ b_{(x,x+1), b}^\dagger b_{(x,x+1), c} + b_{(x,x+1) , d}^\dagger b_{(x,x+1), b} }
	,
\end{equation}
\begin{equation}
	S _ {  x , x + 1 } ^ { 3 } = b_{(x,x+1) , c}^\dagger b_{(x,x+1), c} - b_{(x,x+1) , d}^\dagger b_{(x,x+1), d}
	.
\end{equation}

The energy conditions are listed in Table~\ref{table:energy_conditions_S=1}. Similar to $S = 1/2$, $\Delta_{1, n}$, $\Delta_{2, n}$, $\delta_{1, n}$, $\delta_{2, n}$ for any $n$ are at the order of $\Delta$ mentioned in the main text and are specified in numerical simulations.
\onecolumngrid
\begin{center}
\begin{table}[t]
\begin{center}
\begingroup
\setlength{\tabcolsep}{6pt}
\renewcommand{\arraystretch}{2}
\begin{tabular}{c c}
	\hline\hline
	Molecule State Energy & Value \\
	\hline
	$\epsilon_{S_{2n+1}, a}$ & $ - m - \Sigma_{S_{2n+1}, a; L_{2n+1}, b}$ \\
	$\epsilon_{S_{2n+1} c}$ & $2 h B + \delta_{1, n}$ \\
	\hline
	$\epsilon_{L_{2n+1}, a}$ & $0$ \\
	$\epsilon_{L_{2n+1}, +1}$ ($\epsilon_{L_{2n+1}, c}$) & \begingroup \renewcommand{\arraystretch}{1} \begin{tabular}{@{}c@{}}$ 2 h B + \Delta_{1, n} - \paren{ \delta_{2, n} - \delta_{1, n} } + g^2 / 2 $ \\ $+ \Sigma_{S_{2n+1}, a; L_{2n+1}, d} - \Sigma_{S_{2n+1}, a; L_{2n+1}, b} - \Sigma_{S_{2n+2}, a; L_{2n+2}, d} + \Sigma_{S_{2n+2}, a; L_{2n+2}, b}$\end{tabular} \endgroup \\
	$\epsilon_{L_{2n+1}, 0}$ ($\epsilon_{L_{2n+1}, b}$) & $2 h B + \Delta_{1, n} $ \\
	$\epsilon_{L_{2n+1}, -1}$ ($\epsilon_{L_{2n+1}, d}$) & $2 h B + \Delta_{1, n} + \paren{ \delta_{2, n} - \delta_{1, n} } + g^2 / 2$ \\
	 \hline
	$\epsilon_{S_{2n+2}, a}$ & $m$ \\
	$\epsilon_{S_{2n+2}, c}$ & $2 h B + \delta_{2, n}$ \\
	\hline
	$\epsilon_{L_{2n+2}, a}$ & $\Sigma_{S_{2n+2}, a; L_{2n+2}, b}$ \\
	$\epsilon_{L_{2n+2}, +1}$ ($\epsilon_{L_{2n+2}, c}$) & $2 h B + \Delta_{2, n} - \paren{ \delta_{1, n + 1} - \delta_{2, n} } + g^2 / 2 + \Sigma_{S_{2n+2}, a; L_{2n+2}, d} - \Sigma_{S_{2n+2}, a; L_{2n+2}, b}$ \\
	$\epsilon_{L_{2n+2}, 0}$ ($\epsilon_{L_{2n+2}, b}$) & $2 h B + \Delta_{2, n} $
	 \\
	 $\epsilon_{L_{2n+2}, -1}$ ($\epsilon_{L_{2n+2}, d}$) & $2 h B + \Delta_{2, n} + \paren{ \delta_{1, n + 1} - \delta_{2, n} } + g^2 / 2 - \Sigma_{S_{2n+3}, a; L_{2n+3}, d} + \Sigma_{S_{2n+3}, a; L_{2n+3}, b}$ \\
	 \hline\hline
\end{tabular}
\endgroup
\end{center}
\caption{\label{table:energy_conditions_S=1}Energy conditions for the $n$-th unit cell the $S = 1$ QLM, $n = 0, 1, 2, ...$. The other molecule states not listed are made off resonant. $B$ is the molecule's rotational constant and $h$ is Planck's constant.}
\end{table}
\end{center}

\twocolumngrid
The nearest-neighbor self-interactions with two number operators from the second-order effective Hamiltonian are the total sum of the following terms:
\begin{widetext}
\begin{equation}
\begin{aligned}
	& \Sigma_{S_{2n+1}, a; L_{2n+1}, c} \hat{n}_{S_{2n+1}, a} \hat{n}_{L_{2n+1}, c} + \Sigma_{S_{2n+1}, a; L_{2n+1}, b} \hat{n}_{S_{2n+1}, a} \hat{n}_{L_{2n+1}, b} + \Sigma_{S_{2n+1}, a; L_{2n+1}, d} \hat{n}_{S_{2n+1}, a} \hat{n}_{L_{2n+1}, d} \\
	= & \paren{ \Sigma_{S_{2n+1}, a; L_{2n+1}, c} - 2 \Sigma_{S_{2n+1}, a; L_{2n+1}, b} + \Sigma_{S_{2n+1}, a; L_{2n+1}, d} } \hat{n}_{S_{2n+1}, a} \hat{n}_{L_{2n+1}, c} \\ & + \paren{\Sigma_{S_{2n+1}, a; L_{2n+1}, d} - \Sigma_{S_{2n+1}, a; L_{2n+1}, b}} \paren{- \hat{n}_{L_{2n+1}, c} + \hat{n}_{L_{2n}, d}} + \Sigma_{S_{2n+1}, a; L_{2n+1}, b} \hat{n}_{S_{2n+1}, a}
	,
\end{aligned}
\end{equation}
\begin{equation}
\begin{aligned}
	& \Sigma_{S_{2n+1}, a; L_{2n}, c} \hat{n}_{S_{2n+1}, a} \hat{n}_{L_{2n}, c} + \Sigma_{S_{2n+1}, a; L_{2n}, b} \hat{n}_{S_{2n+1}, a} \hat{n}_{L_{2n}, b} + \Sigma_{S_{2n+1}, a; L_{2n}, d} \hat{n}_{S_{2n+1}, a} \hat{n}_{L_{2n}, d} \\
	= & \paren{ \Sigma_{S_{2n+1}, a; L_{2n}, c} - 2 t_{S_{2n+1}, a; L_{2n}, b} + \Sigma_{S_{2n+1}, a; L_{2n}, d} } \hat{n}_{S_{2n+1}, a} \hat{n}_{L_{2n}, c} \\ & + \paren{\Sigma_{S_{2n+1}, a; L_{2n}, d} - \Sigma_{S_{2n+1}, a; L_{2n}, b}} \paren{ -\hat{n}_{L_{2n+1}, c} + \hat{n}_{L_{2n}, d}} + \Sigma_{S_{2n+1}, a; L_{2n}, b} \hat{n}_{S_{2n+1}, a} 
	,
\end{aligned}
\end{equation}
\begin{equation}
\begin{aligned}
	& \Sigma_{S_{2n+2}, a; L_{2n+1}, c} \hat{n}_{S_{2n+2}, a} \hat{n}_{L_{2n+1}, c} + \Sigma_{S_{2n+2}, a; L_{2n+1}, b} \hat{n}_{S_{2n+2}, a} \hat{n}_{L_{2n+1}, b} + \Sigma_{S_{2n+2}, a; L_{2n+1}, d} \hat{n}_{S_{2n+2}, a} \hat{n}_{L_{2n+1}, d} \\
	= & \paren{ \Sigma_{S_{2n+2}, a; L_{2n+1}, c} - 2 \Sigma_{S_{2n+2}, a; L_{2n+1}, b} + \Sigma_{S_{2n+2}, a; L_{2n+1}, d} } \hat{n}_{S_{2n+2}, a} \hat{n}_{L_{2n+1}, c} \\ & + \paren{\Sigma_{S_{2n+2}, a; L_{2n+1}, d} - \Sigma_{S_{2n+2}, a; L_{2n+1}, b}} \paren{\hat{n}_{L_{2n+1}, d} - \hat{n}_{L_{2n+2}, d}} + \Sigma_{S_{2n+2}, a; L_{2n+1}, b} \hat{n}_{S_{2n+2}, a} 
	,
\end{aligned}
\end{equation}
\begin{equation}
\begin{aligned}
	& \Sigma_{S_{2n+2}, a; L_{2n+2}, c} \hat{n}_{S_{2n+2}, a} \hat{n}_{L_{2n+2}, c} + \Sigma_{S_{2n+2}, a; L_{2n+2}, b} \hat{n}_{S_{2n+2}, a} \hat{n}_{L_{2n+2}, b} + \Sigma_{S_{2n+2}, a; L_{2n+2}, d} \hat{n}_{S_{2n+2}, a} \hat{n}_{L_{2n+2}, d} \\
	= & \paren{ \Sigma_{S_{2n+2}, a; L_{2n+2}, c} - 2 \Sigma_{S_{2n+2}, a; L_{2n+2}, b} + \Sigma_{S_{2n+2}, a; L_{2n+2}, d} } \hat{n}_{S_{2n+2}, a} \hat{n}_{L_{2n+2}, c} \\ & + \paren{\Sigma_{S_{2n+2}, a; L_{2n+2}, d} - \Sigma_{S_{2n+2}, a; L_{2n+2}, b}} \paren{- \hat{n}_{L_{2n+2}, c} + \hat{n}_{L_{2n+1}, c}} + \Sigma_{S_{2n+2}, a; L_{2n+2}, b} \hat{n}_{S_{2n+2}, a}
	,
\end{aligned}
\end{equation}
\end{widetext}
where the Gauss law and other molecular number constraints are already used.

We want the coefficients in front of the two-number operators to vanish, so there are four independent equations for each unit cell generated from this, shown as follows:
\begin{equation}
    \Sigma_{S_{2n+1}, a; L_{2n+1}, c} - 2 \Sigma_{S_{2n+1}, a; L_{2n+1}, b} + \Sigma_{S_{2n+1}, a; L_{2n+1}, d} = 0
    ,
\label{eq:S1_self-interaction_1}
\end{equation}
\begin{equation}
    \Sigma_{S_{2n+1}, a; L_{2n}, c} - 2 \Sigma_{S_{2n+1}, a; L_{2n}, b} + \Sigma_{S_{2n+1}, a; L_{2n}, d} = 0
    ,
\label{eq:S1_self-interaction_2}
\end{equation}
\begin{equation}
    \Sigma_{S_{2n+2}, a; L_{2n+1}, c} - 2 \Sigma_{S_{2n+2}, a; L_{2n+1}, b} + \Sigma_{S_{2n+2}, a; L_{2n+1}, d} = 0
    ,
\label{eq:S1_self-interaction_3}
\end{equation}
\begin{equation}
    \Sigma_{S_{2n+2}, a; L_{2n+2}, c} - 2 \Sigma_{S_{2n+2}, a; L_{2n+2}, b} + \Sigma_{S_{2n+2}, a; L_{2n+2}, d} = 0
    .
\label{eq:S1_self-interaction_4}
\end{equation}
Equations \ref{eq:S1_self-interaction_1}-(\ref{eq:S1_self-interaction_4}) are overdetermined as equations for distances $r$'s, energy parameters $\delta$'s and $\Delta$'s, and the angle $\theta$'s. As is explained in the main text, we have introduced nonequal intermolecular separations such that the self-interactions for molecular pairs $L_{x}$ and $S_{x+1}$ are negligible. Therefore, Eqs.~(\ref{eq:S1_self-interaction_2}) and (\ref{eq:S1_self-interaction_3}) are no longer needed. The remaining equations are underdetermined.  \Cref{eq:S1_self-interaction_1,eq:S1_self-interaction_4} with $\Sigma$'s plugged in are
\begin{widetext}
\begin{equation}
\begin{aligned}
	& \frac{1}{\Delta_{1, n} - \delta_{2, n}} \left| - \frac { 1 } { 4 \pi \epsilon _ { 0 } r _ {S_{2n+1}, L_{2n+1}} ^ { 3 } } \left( 3 \cos ^ { 2 } \theta_ {S_{2n+1}, L_{2n+1}} - 1 \right) \frac{1}{3} d^2 \right|^2 \\
	& - 2 \frac{1}{\Delta_{1, n} - \delta_{1, n}} \left| - \frac { 1 } { 4 \pi \epsilon _ { 0 } r _ {S_{2n+1}, L_{2n+1}} ^ { 3 } } e^{ - i \phi_{S_1, L_1}} \sin {\theta_{S_{2n+1}, L_{2n+1}}} \cos {\theta_{S_{2n+1}, L_{2n+1}}} \frac{1}{\sqrt{2}} d^2 \right|^2 \\
	+ & \frac{1}{\Delta_{1, n} + \delta_{2, n} - 2 \delta_{1, n}} \left| \frac { 1 } { 4 \pi \epsilon _ { 0 } r _ {S_{2n+1}, L_{2n+1}} ^ { 3 } } e^{ - i \phi_{S_{2n+1}, L_{2n+1}}} \sin {\theta_{S_{2n+1}, L_{2n+1}}} \cos {\theta_{S_{2n+1}, L_{2n+1}}} \frac{1}{\sqrt{2}} d^2 \right|^2 = 0
	,
\end{aligned}
\end{equation}
\begin{equation}
\begin{aligned}
& \frac{1}{\Delta_{2, n} - \delta_{1, n + 1}} \left| - \frac { 1 } { 4 \pi \epsilon _ { 0 } r _ {S_2, n; L_2, n} ^ { 3 } } \left( 3 \cos ^ { 2 } \theta_ {S_2, n; L_2, n} - 1 \right) \frac{1}{3} d^2 \right|^2 \\
& - 2 \frac{1}{\Delta_{2, n} - \delta_{2, n}} \left| - \frac { 1 } { 4 \pi \epsilon _ { 0 } r _ {S_2, n; L_2, n} ^ { 3 } } e^{ - i \phi_{S_2, n; L_2, n}} \sin {\theta_{S_2, n; L_2, n}} \cos {\theta_{S_2, n; L_2, n}} \frac{1}{\sqrt{2}} d^2 \right|^2 \\
+ & \frac{1}{\Delta_{2, n} + \delta_{1, n + 1} - 2 \delta_{2, n}} \left| \frac { 1 } { 4 \pi \epsilon _ { 0 } r _ {S_2, n; L_2, n} ^ { 3 } } e^{ - i \phi_{S_2, n; L_2, n}} \sin {\theta_{S_2, n; L_2, n}} \cos {\theta_{S_2, n; L_2, n}} \frac{1}{\sqrt{2}} d^2 \right|^2 = 0 ,
\end{aligned}
\end{equation}
\end{widetext}
After a bit of algebra from the equations, we obtain the constraints
\begin{equation}
	\Delta_{1, n} = \frac{1}{2} \paren{ 3 \delta_{1, n} - \delta_{2, n} }
	,
	\label{eq: Delta_condition_1}
\end{equation}
\begin{equation}
	\Delta_{2, n} = \frac{1}{2} \paren{ 3 \delta_{2, n} - \delta_{1, n + 1} }
	.
	\label{eq: Delta_condition_2}
\end{equation}
From further calculation, there are constraints on angles
\begin{equation}
\begin{aligned}
	& \frac{1}{\sqrt{2}} \left| \sin{\theta_{S_{2n+1}, L_{2n+1}}} \cos{\theta_{S_{2n+1}, L_{2n+1}}} \right| \\ = & \frac{1}{3} \left| \frac{3 \cos^2 \theta_{S_{2n+1}, L_{2n+1}} - 1}{3} \right|
	,
	\label{eq:angle_condition1}
\end{aligned}
\end{equation}
\begin{equation}
\begin{aligned}
	& \frac{1}{\sqrt{2}} \left| \sin{\theta_{S_{2n+2}, L_{2n+2}}} \cos{\theta_{S_{2n+2}, L_{2n+2}}} \right| \\
	= & \frac{1}{3} \left| \frac{3 \cos^2 \theta_{S_{2n+2}, L_{2n+2}} - 1}{3} \right|
	,
	\label{eq:angle_condition2}
\end{aligned}
\end{equation}
which gives
\begin{equation}
    \cos^2 \theta_{S_{2n+1}, L_{2n+1}} = 0.0220216\mathrm{~or~}0.917372
    ,
    \label{eq:sm_theta_1}
\end{equation}
\begin{equation}
    \cos^2 \theta_{S_{2n+2}, L_{2n+2}} = 0.0220216\mathrm{~or~}0.917372
    .
    \label{eq:sm_theta_2}
\end{equation}
In addition to the solutions above, we also need to impose the condition that tunneling amplitudes $-\sqrt{2} w$ at all positions should be equal, which yields some constraints on the distances $r$ and angles $\theta$. In principle, the chain can be zigzag. However, in our scheme, we specifically set the chain of molecules to be a straight line. Namely, all of the angles $\theta$ are the same.

\section{Details on numerical methods and simulation}

We have implemented the exact diagonalization (ED) method to simulate the time evolution of QLMs and the DMH for three unit cells. This section provides the details for the numerical algorithm and the choices of parameters in the simulation.

\subsection{Construction of the Hilbert spaces}

For QLMs, each unit cell has two sites and links $S_{2n + 1}$, $L_{2n + 1}$, $S_{2n + 2}$, and $L_{2n + 2}$ where each site has two degrees of freedom and each link has $2S+1$ degrees of freedom for spin-$S$. Therefore, for $N$ unit cells ($n = 0, 1,..., N - 1$), the Hilbert space should have the dimension $(4S+2)^{2N}$. For the DMH, each unit cell has four molecules and each molecule has four degrees of freedom. It follows that for $N$ unit cells, the dimension of the Hilbert space is $2^{8N}$. It is clear that the DMH Hilbert space dimension is the bottleneck of the ED method when $N$ is large.

To implement the ED method, we need to reduce the dimensionality of both the DMH and QLM Hilbert spaces while preserving the accuracy of the simulation. To achieve that, we utilize symmetries and quantum numbers in both the DMH and the QLM Hamiltonian.
Because of the dipole selection rules and the large value of $B$, the number of molecules at the state $a$ is exactly conserved and thus a good quantum number.
In addition, we can make certain states off resonant by tuning the laser light. For $S=1/2$, states $c$, $d$ on fermion sites and the state $c$ on link sites are off resonant while for $S=1$, states $b$, $d$ on fermion sites are off-resonant. Therefore, those states will be excluded when we construct the DMH Hilbert space.
Since the last link $L_{2 N}$ never changes its state in QLMs due to open boundary conditions, we fix the state of the molecule which represents $L_{2 N}$ to its initial state throughout the simulation for DMH.
For the QLM Hamiltonian, the total number of fermions is conserved and we can use this fact to reduce the dimensionality of the QLM Hilbert space. Similar to the DMH, the state of the last link $L_{2 N}$ is fixed.

According to the above construction of the DMH and QLM Hamiltonian, the DMH Hilbert space is larger than the QLM Hilbert space. The QLM wave function can be embedded into the DMH Hilbert space while the DMH wave function needs to be truncated when it is projected onto the QLM Hilbert space, where the mapping is provided in Table~\ref{table:map} in the main text.
In this paper, fidelity is computed with the embedded QLM wave function in the DMH Hilbert space and DMH observables are computed with the DMH wave function which is projected onto the QLM Hilbert space. Since our constructed QLM Hilbert space, onto which the DMH wave function is projected, is larger than the physical QLM Hilbert space, we will be able to see slight Gauss-law violation in DMH simulations. In experiments, to measure the Gauss law, one can postselect the measurement outcomes of the DMH from our constructed QLM Hilbert space. The Gauss-law plots presented in the main text are computed in the same way.

\subsection{Choices of distances and energy conditions}

The assumed minimum molecule spacing in experiments is $0.5 \ \mu\mathrm{m}$. In this section, all of the energies are divided by Planck's constant $h$ and are thus in units of hertz. With the following specified intermolecular distances and energy conditions, hopping parameters are $w = 82.7 \ \mathrm{Hz}$ for $S = 1/2$ and $w = 3.17 \ \mathrm{Hz}$ for $S = 1$. We note that these choices have yielded decent fidelity overlap with the QLM dynamics (excluding the influence of the ``extra'' gauge-invariant terms in the $S=1$ case), and larger hopping energies can be achieved if sources of dephasing, decoherence, or parameter control disorder serve as practical limitations.

Since the energy conditions of $\Delta_{1,n}$ and $\Delta_{2,n}$ for $S = 1/2$ are flexible, 
for a particular realization of energy conditions for $S =1/2$, Eqs.~(\ref{eq: Delta_condition_1}) and (\ref{eq: Delta_condition_2}) for $S = 1$ are also applicable and can be substituted into Table~\ref{table:energy_conditions_S=1/2_01}, which gives rise to Table~\ref{table:energy_simulation_S05}. For $S = 1$, Eqs.~(\ref{eq: Delta_condition_1}) and (\ref{eq: Delta_condition_2}) are substituted into Table~\ref{table:energy_conditions_S=1} to obtain Table~\ref{table:energy_simulation_S1}.

\onecolumngrid
\begin{center}
\begin{table}[b]
\begin{center}
\begingroup
\setlength{\tabcolsep}{6pt}
\renewcommand{\arraystretch}{2}
\begin{tabular}{c c}
	\hline\hline
	Molecule State Energy & Value \\
	\hline
	$\epsilon_{S_{2 n + 1}, a}$ & $- m - \Sigma_{S_{2n+1}, a; L_{2n+1}, b} - \Sigma_{L_{2n+2}, b; S_{2n+3}, a}$ \\
	$\epsilon_{S_{2 n + 1}, b}$ & $2 h B + \delta_{1, n}$ \\
	\hline
	$\epsilon_{L_{2 n + 1}, a}$ & $0$ \\
	$\epsilon_{L_{2 n + 1}, d}$ & \begingroup \renewcommand{\arraystretch}{1} \begin{tabular}{@{}c@{}} $2 h B + \paren{ 3\delta_{1, n}/2 - \delta_{2, n}/2 } + g^2 / 2$ \\ $- \Sigma_{S_{2n+1}, a; L_{2n+1}, d} - \Sigma_{L_{2n+2}, d; S_{2n+3}, a} + \Sigma_{S_{2n+1}, a; L_{2n+1}, b} + \Sigma_{L_{2n+2}, b; S_{2n+3}, a}$ \end{tabular} \endgroup \\
	$\epsilon_{L_{2 n + 1}, b}$ & $2 h B + \paren{ \delta_{2, n}/2 + \delta_{1, n}/2 } + g^2 / 2$ \\
	 \hline
	$\epsilon_{S_{2 n + 2}, a}$ & $m - \Sigma_{L_{2n+1}, b; S_{2n+2}, a} - \Sigma_{S_{2n+2}, a; L_{2n+2}, d}$ \\
	$\epsilon_{S_{2 n + 2}, b}$ & $2 h B + \delta_{2, n}$ \\
	\hline
	$\epsilon_{L_{2 n + 2}, a}$ & $0$ \\
	$\epsilon_{L_{2 n + 2}, d}$ & $2 h B + \paren{ 3\delta_{2, n}/2 - \delta_{1, n+1}/2 } + g^2 / 2$
	 \\
	 $\epsilon_{S_{2 n + 2}, b}$ & $2 h B + \paren{ \delta_{1, n + 1}/2 + \delta_{2, n}/2 } + g^2 / 2$ \\
	 \hline\hline
\end{tabular}
\endgroup
\end{center}
\caption{\label{table:energy_simulation_S05}Energy conditions used in simulations for the $n$-th unit cell in the $S = 1/2$ QLM, $n = 0, 1, 2, ...$. The other molecule states not listed are made off resonant. $B$ is the molecule's rotational constant and $h$ is Planck's constant.}
\end{table}
\end{center}
\twocolumngrid

\onecolumngrid
\begin{center}
\begin{table}[]
\begin{center}
\begingroup
\setlength{\tabcolsep}{6pt}
\renewcommand{\arraystretch}{2}
\begin{tabular}{c c}
	\hline\hline
	Molecule State Energy & Value \\
	\hline
	$\epsilon_{S_{2 n + 1}, a}$ & $-m - \Sigma_{S_{2n+1}, a; L_{2n+1}, b}$ \\
	$\epsilon_{S_{2 n + 1}, c}$ & $2 h B + \delta_{1, n}$ \\
	\hline
	$\epsilon_{L_{2 n + 1}, a}$ & $0$ \\
	$\epsilon_{L_{2 n + 1}, c}$ & \begingroup \renewcommand{\arraystretch}{1} \begin{tabular}{@{}c@{}}$ 2 h B + \paren{ 5\delta_{1, n}/2 - 3\delta_{2, n}/2 } + g^2 / 2$ \\ $+ \Sigma_{S_{2n+1}, a; L_{2n+1}, d} - \Sigma_{S_{2n+1}, a; L_{2n+1}, b} - \Sigma_{S_{2n+2}, a; L_{2n+2}, d} + \Sigma_{S_{2n+2}, a; L_{2n+2}, b}$\end{tabular} \endgroup \\
	$\epsilon_{L_{2 n + 1}, b}$ & $2 h B + \paren{ 3\delta_{1, n}/2 - \delta_{2, n}/2 } $ \\
	$\epsilon_{L_{2 n + 1}, d}$ & $2 h B + \paren{ \delta_{1, n}/2 + \delta_{2, n}/2 } + g^2 / 2$ \\
	 \hline
	$\epsilon_{S_{2 n + 2}, a}$ & $m$ \\
	$\epsilon_{S_{2 n + 1}, c}$ & $2 h B + \delta_{2, n}$ \\
	\hline
	$\epsilon_{S_{2 n + 2}, a}$ & $\Sigma_{S_{2n+2}, a; L_{2n+2}, b}$ \\
	$\epsilon_{S_{2 n + 2}, c}$ & $2 h B + \paren{ 5\delta_{2, n}/2 - 3\delta_{1, n+1}/2 } + g^2 / 2 + \Sigma_{S_{2n+2}, a; L_{2n+2}, d} - \Sigma_{S_{2n+2}, a; L_{2n+2}, b}$ \\
	$\epsilon_{S_{2 n + 2}, b}$ & $2 h B + \paren{ 3\delta_{2, n}/2 - \delta_{1, n+1}/2 } $
	 \\
	 $\epsilon_{S_{2 n + 2}, d}$ & $2 h B + \paren{ \delta_{2, n}/2 + \delta_{1, n}/2 } + g^2 / 2 - \Sigma_{S_{2n+3}, a; L_{2n+3}, d} + \Sigma_{S_{2n+3}, a; L_{2n+3}, b}$ \\
	 \hline\hline
\end{tabular}
\endgroup
\end{center}
\caption{\label{table:energy_simulation_S1} Energy conditions used in simulations for the $n$-th unit cell the $S = 1$ QLM, $n = 0, 1, 2, ...$. The other molecule states not listed are made off resonant. $B$ is the molecule's rotational constant and $h$ is Planck's constant.}
\end{table}
\end{center}
\twocolumngrid

For both $S = 1/2$ and $S = 1$, we construct two arithmetic sequences $\delta_{1,n}$, $\delta_{2,n}$ by $\delta_{2,n} =  \delta_{1,n} - D_1$ and $\delta_{1,n+1} = \delta_{2,n} - D_2$ to avoid accidental energy degeneracy between two states in different unit cells and thus suppress the first-order interactions in the effective Hamiltonian. We define $V_0 \equiv \frac { 1 } { 4 \pi \epsilon _ { 0 } r _ {S_1, L_1} ^ { 3 } } d^2$, where $d$ is the electric dipole moment of one molecule.
For $S = 1 / 2$, we choose $\delta_{1,0}=25 V_0$, $D_1=20 V_0$, $D_2=140 V_0$, and $B=1000 V_0$. For $S = 1$, we use $\delta_{1,0}=12.5 V_0$, $D_1=10 V_0$, $D_2=70 V_0$, and $B=1000 V_0$. As explained earlier, the experimental value of $B$ should be much higher than any other relevant energy scales in the experiment, so the number of $\ket{a}$ states is an effectively conserved quantity. With this conservation, the value of $B=1000 V_0$ used in the numerical simulation will produce the same result as greater $B$ values.

For $S = 1/2$, in the main text,
\begin{equation}
     - w = \frac { 1 } { 2 } V_{S_x, L_x}^{b, a; a, d} V_{L_x, S_{x+1}}^{b, a; a, b} \left[ \frac{1}{\Delta \epsilon_{1,x}} + \frac{1}{\Delta \epsilon_{2,x}} \right] \ ,
\label{eq:sm_hopping_1}
\end{equation}
with $\Delta \epsilon_{1,x} = \epsilon_{S_x, a} + \epsilon_{L_x, d} - \epsilon_{S_x, b} - \epsilon_{L_x, a}$ and $\Delta \epsilon_{2,x} = \epsilon_{L_x, b} + \epsilon_{S_{x+1}, a} - \epsilon_{L_x, a} - \epsilon_{S_{x+1}, b}$.
$\Delta \epsilon_{1,x}$ and $\Delta \epsilon_{2,x}$ have been specified by energy conditions. The right-hand sides of Eq.~(\ref{eq:sm_hopping_1}) at different $x$'s are required to be the same because the hopping parameter $w$ does not depend on $x$. In order to achieve that, we need to alter the dipole-dipole interactions by tuning the intermolecular distances. Similar tuning needs to be done for $S = 1$ as well for the same reason.
In particular, for both $S = 1/2$ and $S = 1$, we set the relative distance ratios as $r_{S_{2n+1}, L_{2n+1}}=r_{S_1, L_1}$, $r_{L_{2n+1}, S_{2n+2}} = \gamma r_{S_1, L_1}$, $r_{S_{2n+2}, L_{2n+2}}= \beta r_{S_1, L_1}$, $r_{L_{2n+2}; S_{2n+3}} = \beta \gamma r_{S_1, L_1}$ for every $n$, where $\beta = (D_1 / D_2)^{1/6} \approx 0.723$ is fixed by energy conditions and $\gamma$ is defined as a variable long-short distance ratio greater than or equal to one, mentioned in the main text.
$r_{S_{2n+2}, L_{2n+2}}$ are the smallest intermolecular distances, which we assume can be set to $0.5 \ \mu\mathrm{m}$ in an envisioned experiment which corresponds to $r_{S_1, L_1} = 0.692 \ \mu\mathrm{m}$. The $\theta$ angles for all positions are set to be the same with all the molecules on a line. For $S=1/2$, we choose $\cos \theta=0$. For $S = 1$, from Eqs.(\ref{eq:sm_theta_1}) and (\ref{eq:sm_theta_2}), we choose $\cos \theta=0.14840$. The $\phi$ angles are all the same with a line of molecules and are set to zero for both $S = 1/2$ and $S = 1$. They are furthermore irrelevant for one-dimensional models with only local hopping terms. The body-frame electric dipole moment of the ground-state NaRb molecule is $d=3.3 \ \mathrm{D}$ (debye)~\cite{Guo-NaRb}. In the considered experimental setup, $V_0 = 4.96 \ \mathrm{kHz}$ when $r _ {S_1, L_1}=0.692 \ \mu\mathrm{m}$.

In the following, we compute the hopping parameter $w$ with the above energy conditions and with experimentally reasonable parameters. For $S=1/2$, we set $\gamma = 1$. Since $w$ is the same for every position, we can just compute its value at one position,
\begin{equation}
\begin{aligned}
     - w
    = & \left(\frac{\delta_{1,n} - \delta_{2,n}}{2}\right)^{-1} V^{b, a; a, d}_{S_1 L_1} V^{b, a; a, b}_{L_1 S_2} \\
    = & \frac{1}{\delta_{1,n} - \delta_{2,n}} V_0^2 \sin^2 \theta \frac{3 \cos^2 \theta - 1}{6} 
    ,
\end{aligned}
\end{equation}
where as described we have chosen $\delta_{1,n} - \delta_{2,n} = D_1$. From our chosen parameters, it follows that $w = V_0 / 120 = \SI{41.3}{\hertz}$.
For $S=1$, we calculate
\begin{equation}
\begin{aligned}
    & - \sqrt{2} w \\
    = & \left( \frac{3 \paren{\delta_{1,n} - \delta_{2,n}}}{2} \right)^{-1} \left( -\frac{1}{ 4 \pi \epsilon_{0} r_{S_1, L_1}^{3}} d^2  \frac{3 \cos^2 \theta - 1}{3}\right) \\ & \times \left(-\frac{1}{ 4 \pi \epsilon_{0} r_{L_1; S_2}^{3}} d^2 \frac{1}{\sqrt{2}} \sin \theta \cos \theta \right) \\
    = & - \left( \frac{3 D_1}{2} \right)^{-1} \frac{V_0^2}{\gamma^3} \frac{1}{3 \sqrt{2}} \left( 3 \cos^2 \theta - 1 \right) \sin \theta \cos \theta \\
    = & 0.000638 V_0
    .
	\label{eq:sm_hopping_S1}
\end{aligned}
\end{equation}
We choose the long-short distance ratio $\gamma = 1.5$ in the main text. From our chosen parameters, it follows that $\sqrt{2} w = \SI{3.17}{\hertz}$. We note that this rather small energy scale may be practically challenged by both dephasing and parameter control disorder, and larger values can be achieved by relaxing some of the assumed energy constraint conditions.

\subsection{Symmetries exhibited by densities in QLM simulations}

Here we remark that there are certain symmetries and conserved quantities explicitly exhibited in QLM simulations starting with our prepared initial states. For the $S=1/2$ QLM, the dynamic of the density on each site or link is invariant under the change of mass from $m$ to $-m$. In QLMs with open boundary conditions for any $S$, there exists a $CP$ symmetry manifested by dynamical densities as well.

The Hamiltonian of the $S=1/2$ QLM is
\begin{equation}
\begin{aligned}
    H = & H_{\mathrm{hopping}} + H_{\mathrm{mass}} \\
    = & - w \sum _ { x } \left[ \psi _ { x } ^ { \dagger } U _ { x , x + 1 } \psi _ { x + 1 } + \text{H.c.} \right] + m \sum _ { x } ( - 1 ) ^ { x } \psi _ { x } ^ { \dagger } \psi _ { x }
    ,
\end{aligned}
\end{equation}
where $H_{\mathrm{hopping}}$ is the fermion hopping term, $H_{\mathrm{mass}}$ is the fermion mass term, and we have discarded the constant electric flux energy term. The Hamiltonian with an opposite mass term is denoted as
\begin{equation}
\begin{aligned}
    & H' \\ = & H_{\mathrm{hopping}} - H_{\mathrm{mass}} \\ = & - w \sum _ { x } \left[ \psi _ { x } ^ { \dagger } U _ { x , x + 1 } \psi _ { x + 1 } + \text{H.c.} \right] - m \sum _ { x } ( - 1 ) ^ { x } \psi _ { x } ^ { \dagger } \psi _ { x }
    .
\end{aligned}
\end{equation}

In discussing the inversion symmetry of the mass term for $S = 1/2$, we need to pick a specific basis of the QLM Hilbert space in order to implement an explicit complex conjugation which is basis dependent. After a specific basis is chosen, we will just work with vectors and matrices comprised of complex numbers instead of Dirac's bras and kets. A natural basis to use is comprised of tensor products of single site and/or link states
\begin{equation}
    \ket{f_1}_{S_1} \ket{E_{1,2}}_{L_1} \ket{f_2}_{S_2} \ket{E_{2,3}}_{L_2} \cdots
    ,
\end{equation}
where the fermion occupation number $f_x$ takes values of $f_x = 1$ for ``occupied'' or $f_x = 0$ for ``unoccupied'' and $E_{x, x+1}$ is the electric flux on the link between sites $x$ and $x+1$. In this basis, the matrix elements of the QLM Hamiltonian $H$ are all real numbers because the coefficients $-w$ and $m$ are real.

We consider a real column vector (all of the components of which are real) $\psi_0$ in our chosen basis as the initial state of real-time evolution. The real-time evolution of our initial state is $e^{- i H t} \psi_0$, where $H$ refers to the aforementioned real matrix. The expectation value of a Hermitian quantum operator with the Hamiltonian $H$ as a function of time is
\begin{equation}
   \langle O (t) \rangle_H \equiv \psi_0^\dagger e^{i H t} O e^{- i H t} \psi_0
   ,
\end{equation}
where $O$ is the representation of the quantum operator in our chosen basis. The expectation value of the same operator with the other Hamiltonian $H'$ is
\begin{equation}
   \langle O (t) \rangle_{H'} \equiv \psi_0^\dagger e^{i H' t} O e^{- i H' t} \psi_0
   .
\end{equation}
We want to figure out the conditions for $\langle O (t) \rangle_H = \langle O (t) \rangle_{H'}$ at any time $t$. It will be done in two steps, taking the complex conjugate such that $H' = H_{\mathrm{hopping}} - H_{\mathrm{mass}} \rightarrow - H' = - H_{\mathrm{hopping}} + H_{\mathrm{mass}}$ and carrying out a diagonal unitary transformation such that $- H' = - H_{\mathrm{hopping}} + H_{\mathrm{mass}} \rightarrow  H = H_{\mathrm{hopping}} + H_{\mathrm{mass}}$.

$\langle O (t) \rangle_{H'}$ is real because $O$ is Hermitian, and thus we can write
\begin{equation}
\begin{aligned}
    & \langle O (t) \rangle_{H'} \\
    = & \langle O (t) \rangle_{H'}^* \\
    = & \psi_0^T e^{ - i H' t} O e^{ i H' t} \psi_0^* \\
    = & \psi_0^\dagger e^{ - i H' t} O e^{ i H' t} \psi_0
    ,
\end{aligned}
\label{eq:expectation_val_O_1}
\end{equation}
where in the last line we have used the fact that all of the components of $\psi_0$ are real. Then we consider to implement a diagonal unitary transformation $U$ whose diagonal matrix elements are given by
\begin{widetext}
\begin{equation}
\begin{aligned}
    & \left( \cdots \bra{f_x}_{S_x} \bra{E_{x,x+1}}_{L_x} \bra{f_{x+1}}_{S_{x+1}} \bra{E_{x+1,x+2}}_{L_{x+1}} \cdots \right) U \left( \cdots \ket{f_x}_{S_x} \ket{E_{x,x+1}}_{L_x} \ket{f_{x+1}}_{S_{x+1}} \ket{E_{x+1,x+2}}_{L_{x+1}} \cdots \right) \\
    = & (-1)^{\sum_{x~\mathrm{odd}} f_x} 
    ,
\end{aligned}
\end{equation}
\end{widetext}
where the bra and the ket only differ in the state of a site $S_x$. All of the off-diagonal matrix elements of $U$ vanish. $U$ takes $\psi_x$ to $- \psi_x$ for any odd $x$, i.e., $U \psi_x U = - \psi_x$. Similarly, $U \psi_x^\dagger U = - \psi_x^\dagger$ for any odd $x$. Therefore, we have $U H_{\mathrm{hopping}} U^\dagger = - H_{\mathrm{hopping}}$ and $U H_{\mathrm{mass}} U^\dagger = H_{\mathrm{mass}}$. $U$ acting on $\psi_0$ gives a global phase $U \psi_0 = e^{i \varphi_0} \psi_0$ where $\varphi_0 = \pi \sum_{x~\mathrm{odd}} f_x$ depends on details of $\psi_0$. Following Eq. (\ref{eq:expectation_val_O_1}), we further derive
\begin{equation}
\begin{aligned}
    & \langle O (t) \rangle_{H'} \\
    = & \psi_0^\dagger e^{ - i H' t} O e^{ i H' t} \psi_0 \\
    = & \psi_0^\dagger U^\dagger \left( U e^{ - i H' t} U^\dagger \right) \left( U O U^\dagger \right) \left( U e^{ i H' t} U^\dagger \right)  U \psi_0 \\
    = & \psi_0^\dagger e^{- i \varphi_0} e^{ i H t} \left( U O U^\dagger \right) e^{ - i H t} e^{i \varphi_0} \psi_0 \\
    = & \psi_0^\dagger e^{ i H t} \left( U O U^\dagger \right) e^{ - i H t} \psi_0
    .
\end{aligned}
\label{eq:expectation_val_O_2}
\end{equation}
Clearly, if $U O U^\dagger = O$, then we will arrive at $\langle O (t) \rangle_H = \langle O (t) \rangle_{H'}$. The condition $U O U^\dagger = O$ is true for any densities of sites or links because density operators are diagonal in our chosen basis. This accounts for why the densities in our numerical studies exhibit an invariance under $m \rightarrow -m$.

$S = 1$ QLMs do not have the inversion symmetry of the mass term.

The parity $P$ and the charge conjugation $C$ transformations for QLM operators are given by \cite{Banerjee-GaugeFields-12}
\begin{equation}
    P^{-1} \psi_x P  = \psi_{-x}
    ,
\end{equation}
\begin{equation}
    P^{-1} \psi_x^\dagger P  = \psi_{-x}^\dagger
    ,
\end{equation}
\begin{equation}
    P^{-1} U_{x, x+1} P  = U_{-x-1, -x}^\dagger
    ,
\end{equation}
\begin{equation}
    P^{-1} E_{x, x+1} P  = - E_{-x-1, -x}
    ,
\end{equation}
\begin{equation}
    C^{-1} \psi_x C  = \left( -1 \right)^{x+1} \psi_{x+1}^\dagger
    ,
\end{equation}
\begin{equation}
    C^{-1} \psi_x^\dagger C  = \left( -1 \right)^{x+1} \psi_{x+1}
    ,
\end{equation}
\begin{equation}
    C^{-1} U_{x, x+1} C  = U_{x+1, x+2}^\dagger
    ,
\end{equation}
\begin{equation}
    C^{-1} E_{x, x+1} C  = - E_{x+1, x+2}
    .
\end{equation}
The $CP$ transformation (a $P$ transformation followed by a $C$ transformation) on QLM operators is
\begin{equation}
    C^{-1} P^{-1} \psi_x P C = \left( -1 \right)^{-x+1} \psi_{-x+1}^\dagger
    ,
\end{equation}
\begin{equation}
    C^{-1} P^{-1} \psi_x^\dagger P C = \left( -1 \right)^{-x+1} \psi_{-x+1}
    ,
\end{equation}
\begin{equation}
    C^{-1} P^{-1} U_{x, x+1} P C = U_{-x, -x+1}
    ,
\end{equation}
\begin{equation}
    C^{-1} P^{-1} E_{x, x+1} P C = E_{-x, -x+1}
    .
\end{equation}
It can be checked that the QLM Hamiltonian $H$ is $CP$ symmetric, i.e., $C^{-1} P^{-1} H P C = H$.

We then show that there are observables exhibiting the $CP$ symmetry in time evolution of a $CP$-symmetric initial state. Such an initial state $\ket{\psi (0)} = \ket{\psi_0}$ obeys
\begin{equation}
    P C \ket{\psi_0} = e^{i \varphi_0} \ket{\psi_0}
    ,
\end{equation}
where $\varphi_0$ is a constant phase with no physical meaning in itself and may depend on $\ket{\psi_0}$. We consider a $CP$-odd operator that by definition, obeys
\begin{equation}
    C^{-1} P^{-1} O P C = - O
    .
\end{equation}
The expectation value of $O$ evaluated at the time evolution of $\ket{\psi_0}$ is
\begin{widetext}
\begin{equation}
\begin{aligned}
    & \mel{\psi_0}{e^{i H t} O e^{-i H t}}{\psi_0} \\
    = & \left( \bra{\psi_0}C^{-1} P^{-1} \right) \left( P C e^{i H t} C^{-1} P^{-1} \right) \left( P C O C^{-1} P^{-1} \right) \left( P C e^{-i H t} C^{-1} P^{-1} \right) \left( P C \ket{\psi_0} \right) \\
    = & \mel{\psi_0}{ e^{-i \varphi_0} e^{i H t} \left( - O \right) e^{-i H t} e^{i \varphi_0} }{\psi_0} \\
    = & - \mel{\psi_0}{e^{i H t} O e^{-i H t}}{\psi_0} \\
    = & 0
    .
\end{aligned}
\end{equation}
\end{widetext}
It can be checked that $\psi_x^\dagger \psi_x + \psi_{-x+1}^\dagger \psi_{-x+1} - 1$ and $E_{x, x+1} - E_{-x, -x+1}$ are both $CP$-odd operators.

\begin{figure}[hbt!]
  \centering
  \includegraphics[width=1.0\columnwidth]{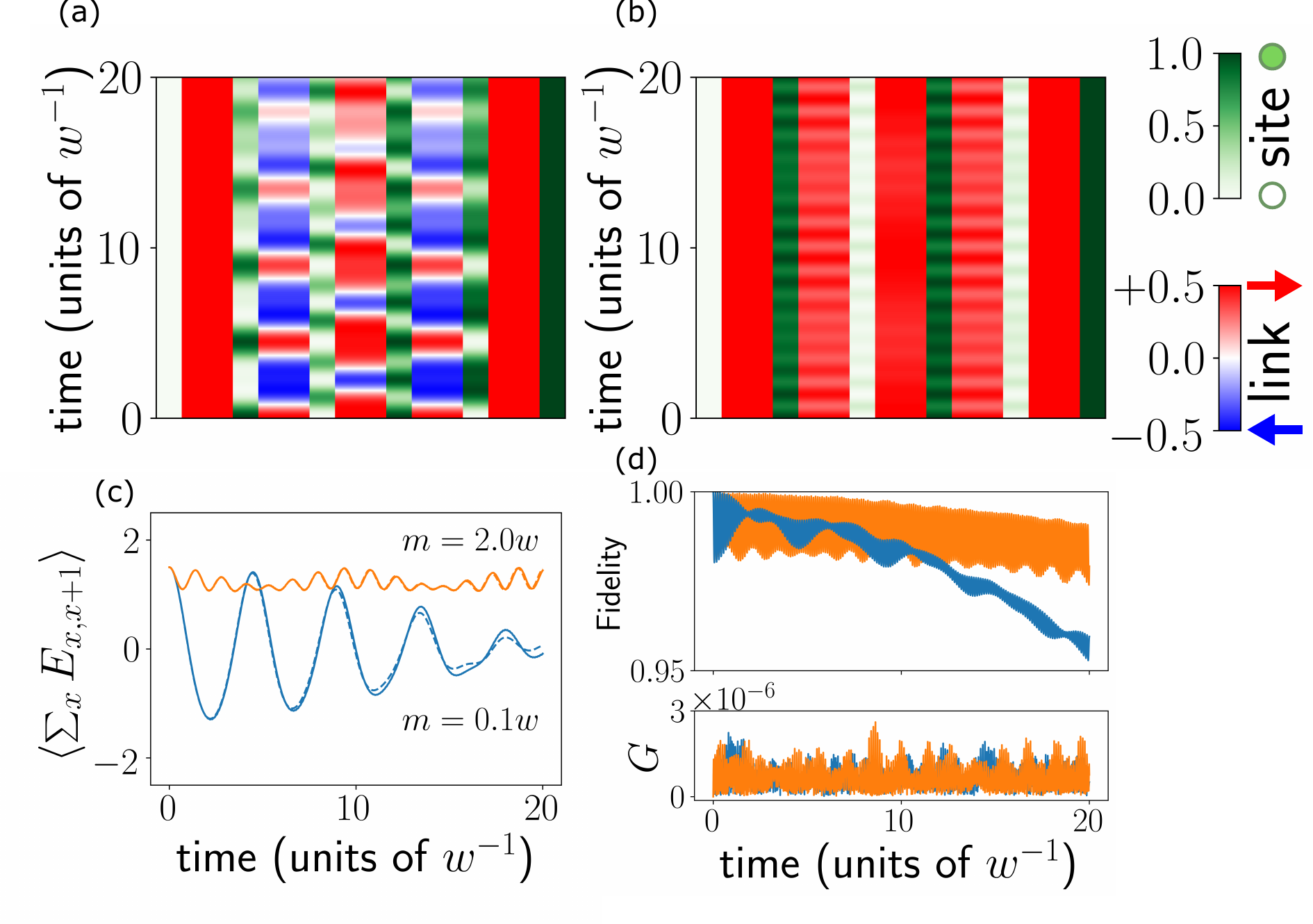}
  \caption{
  Real-time evolution of densities of sites and links in the dipolar molecular system to simulate the $S=1/2$ QLM on two unit cells. The dynamics for initialized strings of right-pointing electric fields are shown for the cases of small mass, (a) $m = 0.1 w$, and large mass, (b) $m = 2.0 w$. Time is in units of the inverse hopping, $w^{-1}$. For small mass (a), the electric field of the string undergoes large-scale oscillations. For large mass (b), the string stays roughly fixed, with only small fluctuations of the charge densities and link spins.
  To note for both (a) and (b), the outermost sites and links are fixed because of the open boundary conditions.
  (c)~Electric fluxes summed over all dynamical links for $m = 0.1 w$ (blue) and $m = 2.0 w$ (orange). Solid and dashed lines relate to the \hyperref[eq:dipole-mol-ham]{DMH} and \hyperref[eq:QLM]{QLM} dynamics, respectively.
  (d) Top: Fidelity of the dipolar molecular wavefunction versus the QLM wavefunction. 
  Bottom: The effective gauge-invariance parameter $G \equiv \sum_x | \langle  \tilde{G}_x \rangle | / L$ \cite{Banerjee-GaugeFields-12} at the two mass values shown in (a) and (b).
  }
  \label{fig:four_panel_2uc_S05}
  \end{figure}
  \begin{figure}
    \includegraphics[width=1.0\columnwidth]{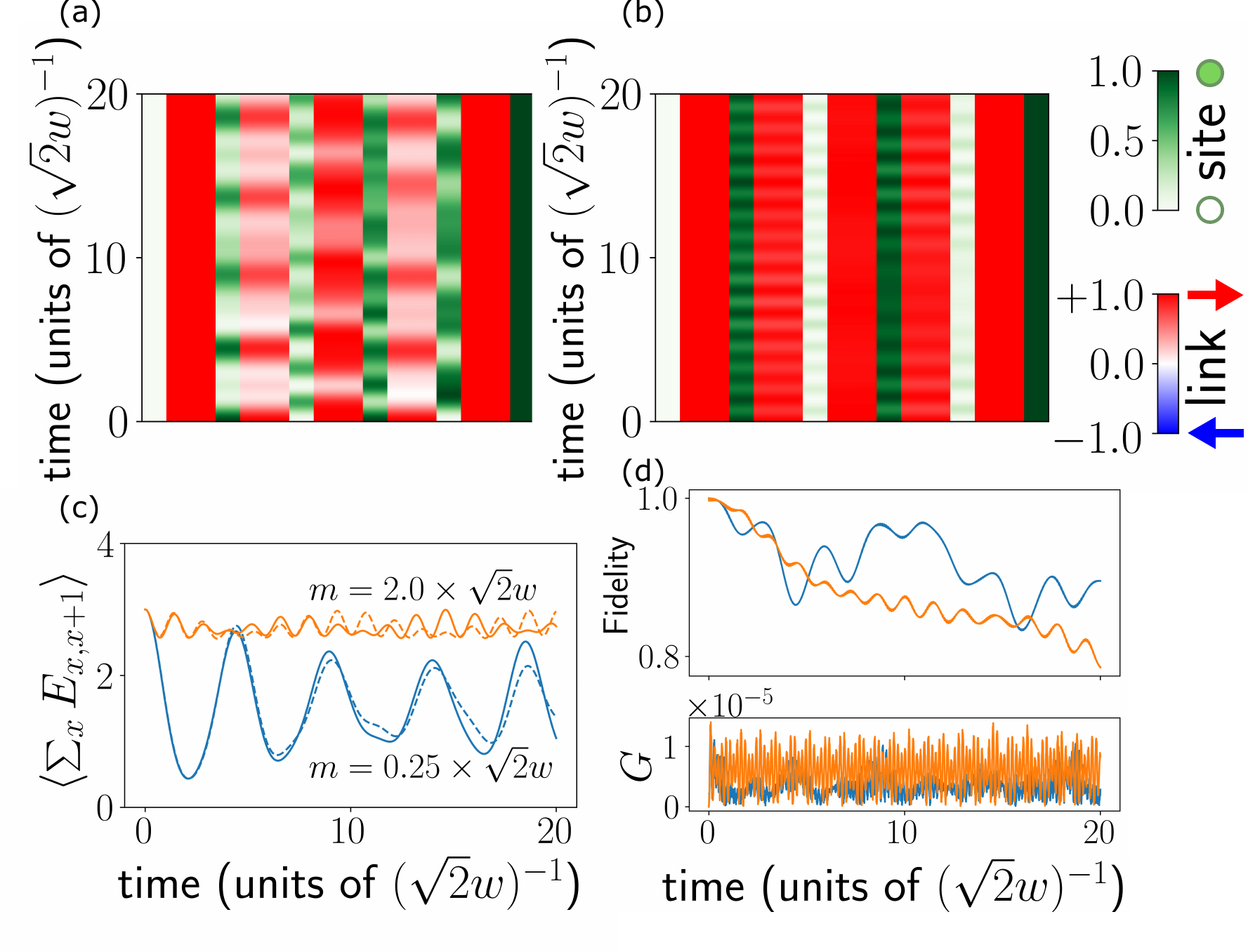}
  \caption{
  Real-time evolution of fermions and links in the dipolar molecular system to simulate the $S=1$ QLM on two unit cells with $g^2 = \sqrt{2} w$ starting from a string of right-pointing electric fields at (a) $m = 0.25 \times \sqrt{2} w$ and (b) $m = 2.0 \times \sqrt{2} w$. Time is in units of $(\sqrt{2} w)^{-1}$. With a small mass (a), the string breaks (modulo finite size effects~\cite{Pichler-PRX}) reaching values close to zero on the hopping timescale, resulting in two approximate mesons on the edges and approximate vacuum in between. With a large mass (b), the string approximately remains, with small fluctuations in densities.
  In (a) and (b), the densities of the two sites and two links on the edges are fixed due to the open boundary condition.
  (c)~Electric fluxes summed over all dynamical links for both $m = 0.25 \times \sqrt{2} w$ (blue) and $m = 2.0 \times \sqrt{2} w$ (orange). Solid and dashed lines relate to the \hyperref[eq:dipole-mol-ham]{DMH} and \hyperref[eq:QLM]{QLM} dynamics, respectively. (d)~Top: Fidelity of the dipolar molecular wavefunction versus the QLM wavefunction.
  Bottom: The gauge-invariance invariance parameter $G \equiv \sum_x | \langle  \tilde{G}_x \rangle | / L$~\cite{Banerjee-GaugeFields-12} at two values of masses in (a) and (b).
  }
  \label{fig:four_panel_2uc_S1}
\end{figure}

In our envisioned systems with open boundary conditions with $N$ unit cells, we can relabel the sites using indices $-\textit{N+1}, -\textit{N+2}, ..., 0, 1, ..., \textit{N}$, consistent with indices used in the above $C P$ transformation. When the initial state is prepared as a $CP$-symmetric state as in the main text, expectation values of $\psi_x^\dagger \psi_x + \psi_{-x+1}^\dagger \psi_{-x+1} - 1$ and $E_{x, x+1} - E_{-x, -x+1}$ for any valid $x$ always vanish at any time $t$. The time evolution of the $CP$-symmetric initial state is also $CP$ symmetric and this fact is manifested by density expectation values, as is illustrated in numerical results in the main text.

\section{Comparison with the two-unit-cell case}

We provide the two-unit-cell results in Figs.~\ref{fig:four_panel_2uc_S05} and \ref{fig:four_panel_2uc_S1} for comparing with the three-unit-cell results given in the main text, using the same parameters. This comparison is to give a glimpse of the scaling with the system size, within the capability of our ED method. The oscillation periods in the two-unit-cell case are slightly greater than the three-unit-cell case. The fidelity in the two-unit-cell case out to $t = 20 w^{-1}$ for $S=1/2$ or $t = 20 
(\sqrt{2} w)^{-1}$ is greater than that in the three-unit-cell case. Finite size effects can make dynamics in two unit cells different from that in three unit cells.

\clearpage

\bibliography{main}

\end{document}